\documentclass[useAMS,usenatbib]{mnras}
\usepackage{aasmacros,placeins,float,times}
\usepackage{amsmath}
\usepackage{graphicx,url,pifont}
\usepackage{amssymb,color}

\newcommand{\cmark}{\ding{51}}%
\newcommand{\xmark}{\ding{55}}%
\def\gsim{\mathrel{\hbox{\rlap{\hbox{\lower4pt\hbox{$\sim$}}}\hbox{$>$}}}}

\def\galform~{\textsc{galform}}

\newcommand{\bq}{\begin{eqnarray}}
\newcommand{\eq}{\end{eqnarray}}

\title[Reionsation in sterile neutrino cosmologies]{Reionisation in
  sterile neutrino cosmologies}

\author[S. Bose et al.]{Sownak Bose$^{1}$\thanks{E-mail:
    sownak.bose@durham.ac.uk}, Carlos S. Frenk$^{1}$, Hou Jun$^{1}$,
  Cedric G. Lacey$^{1}$ and Mark R. Lovell$^{2,3}$ \\ $^{1}$Institute
  for Computational Cosmology, Durham University, South Road, Durham,
  UK, DH1 3LE \\ $^{2}$GRAPPA Institute, Universiteit van Amsterdam,
  Science Park 904, 1098 XH Amsterdam, The Netherlands
  \\ $^{3}$Instituut-Lorentz for Theoretical Physics, Niels Bohrweg 2,
  NL-2333 CA Leiden, The Netherlands \\ }
   
\begin{document}

\pagerange{\pageref{firstpage}--\pageref{lastpage}} \pubyear{2016}

\maketitle

\label{firstpage}

\begin{abstract}
 We investigate the process of reionisation in a model in which the
 dark matter is a warm elementary particle such as a sterile
 neutrino. We focus on models that are consistent with the dark
 matter decay interpretation of the recently detected line at 3.5 keV
 in the X-ray spectra of galaxies and clusters. In warm dark matter
 models the primordial spectrum of density perturbations has a cut-off
 on the scale of dwarf galaxies. Structure formation therefore begins
 later than in the standard cold dark matter (CDM) model and very few
 objects form below the cut-off mass scale. To calculate the number of
 ionising photons, we use the Durham semi-analytic model of galaxy
 formation, \galform~.  We find that even the most extreme 7 keV
 sterile neutrino we consider is able to reionise the Universe early
 enough to be compatible with the bounds on the epoch of reionisation
 from {\it Planck}. This, perhaps surprising, result arises from the
 rapid build-up of high redshift galaxies in the sterile neutrino
 models which is also reflected in a faster evolution of their far-UV
 luminosity function between $10>z>7$ than in CDM. The dominant
 sources of ionising photons are systematically more massive in the
 sterile neutrino models than in CDM. As a consistency check on the
 models, we calculate the present-day luminosity function of
 satellites of Milky Way-like galaxies. When the satellites recently
 discovered in the DES survey are taken into account, strong
 constraints are placed on viable sterile neutrino models.
 \end{abstract}
 
\section{Introduction}

Dark matter, the non-baryonic component that makes up the majority of
the mass of the Universe, is the foundation of today's cosmological
paradigm. The standard model, $\Lambda$CDM, assumes that the dark
matter is a cold, collisionless particle and that the energy density
of the Universe today is dominated by dark energy in the form of a
cosmological constant. This model has predictive power and accounts
for basic measurements of the evolution of large-scale structure in
our Universe, from the temperature anisotropies in the cosmic
microwave background radiation (CMB) at early times
\citep{Planck2015}, to the statistics of the galaxy clustering pattern
today (e.g., \citealt{Cole2005,Eisenstein2005,Zehavi2011}). Its main
shortcoming at present is that the cold particles have not yet been
conclusively detected \citep[but see][]{Hooper2011}.

Cold particles are not the only well-motivated candidates for the dark
matter. An example of a different kind of particle is the {\it sterile
  neutrino}
\citep{Dodelson1994,Abazajian2001a,Abazajian2001b,Dolgov2002}, which
appears in a simple extension of the Standard Model. Its interaction
with active neutrinos could source neutrino flavour oscillations. In
order simultaneously to account for the dark matter and flavour
oscillations, {\it at least} three right-handed sterile neutrinos are
needed \citep{Asaka2005,Asaka2007,Canetti2013}. In this `Neutrino
Minimal Standard Model' (or $\nu$MSM), two of the sterile neutrinos
interact more strongly with the active neutrinos than the third, which
behaves as dark matter \citep{Boyarsky2009}. With the appropriate
choice of parameters in the Lagrangian, it is possible to obtain the
correct dark matter density in sterile neutrinos.

Interest in $\nu$MSM has been boosted recently by the detection of an
X-ray line at 3.5~keV in the stacked spectrum of galaxy clusters
\citep{Bulbul2014a}, M31 and the Perseus cluster
\citep{Boyarsky2014a}. According to these authors, the excess at
3.5~keV cannot be explained by any known metal lines and could, in
fact, be the result of the decay of sterile neutrinos with a rest mass
of 7~keV. This interpretation of the line has subsequently been
challenged by several authors (see, for example,
\citealt{RiemerSorensen2014,Malyshev2014,Jeltema2015,Anderson2015}).
Most recently, \cite{Jeltema2015b} failed to detect any excess at
3.5~keV in a deep {\it XMM-Newton} observation of the dwarf spheroidal
galaxy Draco, attributing the original line detection to an excitation
of K~VIII. Crucially, however, the \cite{Jeltema2015b} analysis made
use of only a subset of the data; with the complete dataset and an
alternative model for the backgrounds, \cite{Ruchask2015} detected
positive residuals at 3.5 keV at 2.3$\sigma$ significance, with a flux
consistent with those obtained from the original stacked galaxy
cluster and M31 observations. Future X-ray observatories may establish
the true identity of this line.

From the point of view of cosmology, the defining property of keV mass
sterile neutrinos is that they behave as {\it warm dark matter} (WDM).
In contrast to CDM, warm particles are kinematically energetic at
early times and thus free stream out of small-scale primordial
perturbations, inducing a cut-off in the power spectrum of density
fluctuations. On large scales unaffected by the free streaming
cut-off, structure formation is very similar in CDM and sterile
neutrino cosmologies (and in WDM in general), but on scales comparable
to or smaller than the cut-off, structure formation proceeds in a
fundamentally different way in the two cases. No haloes form below a
certain mass scale determined by the cut-off and the formation of
small haloes above the cut-off is delayed
\citep[see][]{Bode2001,AvilaReese2001,Viel2005,Lovell2012,Schneider2012,Bose2016,Bose2016b}

For a 7~keV sterile neutrino, the cut-off mass is $\sim 10^9 M_\odot$.
Thus, potentially observable differences from CDM would emerge on
subgalactic scales and at high redshifts when the delayed onset of
structure formation might become apparent. The Local Group and the
early Universe are thus good hunting grounds for tell-tale signs that
might distinguish warm from cold dark matter. There is now a wealth of
observational data for small galaxies in the Local Group (e.g.
\citealt{Koposov2008,McConnachie2012}), as well as measurements of the
abundance of galaxies at high redshifts (e.g.
\citealt{uv_lf_mclure,uv_lf_bouwens2015}) and estimates of the
redshift of reionisation \citep{Planck2015}. One might hope that these
data could constrain the parameters of WDM models
\citep[e.g.][]{Schultz2014,Abazajian2014,Calura2014,Dayal2015a,Dayal2015b,Governato2015,Lovell2015,Maio2015,Bozek2016}.

In this work, we address these questions using the Durham
semi-analytic model of galaxy formation, \galform~
\citep{Cole2000,Lacey2015}, applied both to CDM and sterile neutrino
dark matter. The model follows the formation of galaxies in detail
using a Monte Carlo technique for calculating halo merger trees and
well-tested models for the baryon physics that result in the formation
of visible galaxies. \galform~ predicts the properties of the galaxy
population at all times. This approach has the advantage that it can
easily generate large statistical samples of galaxies at high
resolution for a variety of dark matter models which would be
prohibitive in terms of computational time with the current generation
of hydrodynamic simulations.

This paper is structured as follows. In Section~\ref{sterile} we
introduce the concept of sterile neutrinos and the models considered
in this paper. In Section~\ref{galaxy} we describe the astrophysical
motivation behind this work, as well as the semi-analytic model,
\galform~, used in our analysis. Our results are presented in
Section~\ref{results} and our main conclusions summarised in
Section~\ref{conclusions}.

\section{The sterile neutrino model}
\label{sterile}

Sterile neutrinos\footnote{These particles are `sterile' in the sense
  that they do not interact via the weak force, as is the case for
  active neutrinos in the Standard Model.}  are relativistic when they
decouple and therefore have non-negligible velocities which smear out
density perturbations on small scales.  Hence, sterile neutrinos
behave as WDM. In the original model introduced by
\cite{Dodelson1994}, sterile neutrinos are created by non-resonant
mixing with active neutrinos in the Standard Model. The scale of the
free streaming is determined solely by the rest mass of the sterile
neutrino -- the lighter the particle, the larger the free streaming
length, and the larger the scales at which differences relative to CDM
appear.

\cite{Shi1999} proposed an alternative production mechanism in which
the abundance of sterile neutrinos is boosted by a primordial {\it
  lepton asymmetry}. The value of this quantity, which measures the
excess of leptons over anti-leptons, affects the scale of free
streaming in addition to the rest mass of the sterile neutrino.
\cite{Asaka2005} proposed a model for the generation of the lepton
asymmetry by introducing three right-handed sterile neutrinos in what
is known as the `Neutrino Minimal Standard Model' ($\nu$MSM, see also
\citealt{Boyarsky2009}). In this model, a keV mass sterile neutrino
(labelled $N_1$) is partnered with two GeV mass sterile neutrinos
($N_2$ and $N_3$). It is $N_1$ that behaves as the dark matter, with
its keV mass ($M_1$) leading to early free streaming. The decay of
$N_2$ and $N_3$ prior to the production of $N_1$ generates significant
lepton asymmetry; this boosts the production of $N_1$ via resonant
mixing. Here, we formally quantify the lepton asymmetry, or $L_6$, as:
\bq L_6 \equiv 10^6\left( \frac{n_{\nu_e} - n_{\bar{\nu_e}}}{s}
\right)\,, \eq 
where $n_{\nu_e}$ is the number density of electron neutrinos,
$n_{\bar{\nu}_e}$ the number density of electron anti-neutrinos and
$s$ is the entropy density of the Universe \citep{Laine2008}.

A third parameter in the $\nu$MSM is the {\it mixing angle},
$\theta_1$. The requirement that the model should achieve the correct
dark matter abundance for a given sterile neutrino rest mass uniquely
fixes the value of $\theta_1$ for a particular choice of $L_6$. The
X-ray flux, $F$, associated with the decay of $N_1$ is then
proportional to $\sin^2\left(2\theta_1\right)M_1^5$. We refer the
reader to \cite{Venumadhav2015} and \cite{Lovell2015} for a more
comprehensive discussion of the sterile neutrino model.

In this paper we are particularly interested in sterile neutrinos that
could decay to produce two 3.5~keV photons. We therefore fix the mass
$M_1 = 7\, \rm{keV}$. At this mass, the `warmest' and `coldest'
sterile neutrino models that achieve the correct dark matter density
correspond to $L_6=700$ and $L_6=8$ respectively. By this we mean that
the $L_6=700$ model exhibits deviations from CDM at larger mass scales
than the $L_6=8$ model, which produces similar structure to CDM down to
the scale of dwarf galaxies.

For the $L_6=700$ case, however, the corresponding mixing angle (which
we remind the reader is now {\it fixed}) does not lead to the X-ray
decay flux required to account for the observations of
\cite{Bulbul2014a} and \cite{Boyarsky2014a}. For this reason, we
additionally consider the case $L_6=12$, which corresponds to the
warmest 7 keV sterile neutrino model that has the correct dark matter
abundance {\it and} produces the correct flux at 3.5~keV. This
information is summarised in Table~\ref{dmprops}. Here, we also quote
a characteristic wavenumber, $k_{1/4}$, which measures the scale at
which the linear power spectrum for a given $L_6$ has $1/4$ of the
power of the CDM linear power spectrum. This parameter characterises
the `warmth' of the model. The most extreme case ($L_6=700$) has
$k_{1/4} = 16.05~h/\rm{Mpc}$, whereas the model closest to CDM
($L_6=8$) has $k_{1/4} = 44.14~h/\rm{Mpc}$.

\begin{figure} \centering \includegraphics[trim=0.6cm 0.4cm 0.4cm
  0.4cm,clip=true,width=0.45\textwidth]{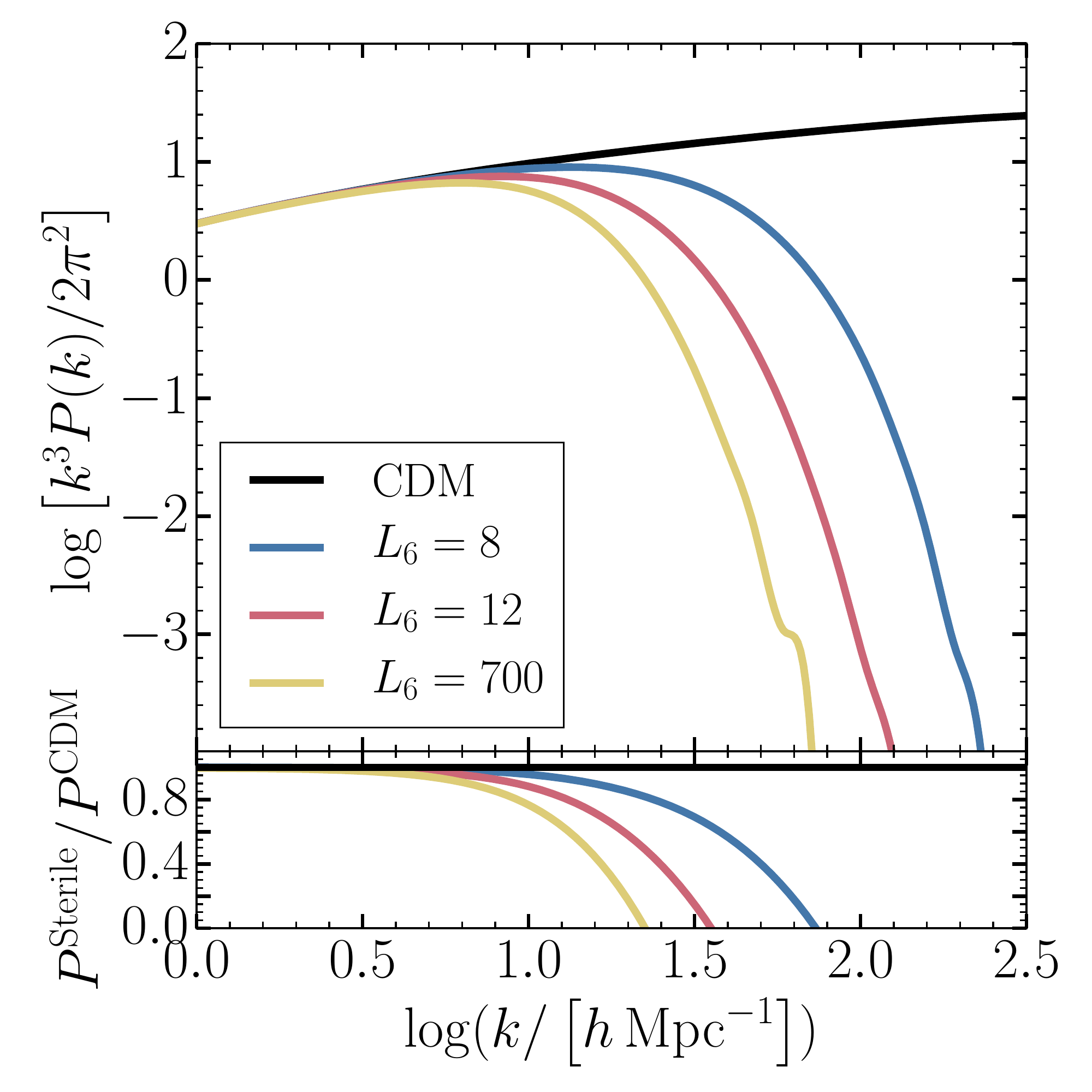}
  \caption{{\em Top panel:} The dimensionless matter power spectra for
    the different dark matter candidates considered in this paper. In
    addition to CDM, we consider a 7 keV sterile neutrino with three
    values of $L_6 = \left(8, 12, 700\right)$, shown with the colours
    indicated in the legend. For the same sterile neutrino mass,
    different $L_6$ values lead to deviations from CDM on different
    scales, with the most extreme case being the $L_6 = 700$
    model. {\em Bottom panel}: The ratio of each power spectrum to
    that of CDM.}
\label{ps_models}
\end{figure}

Fig.~\ref{ps_models} shows the linear power spectrum (in arbitrary
units) of these three models ($L_6=\left(8,12,700\right)$), with the
CDM power spectrum also plotted for comparison. The cosmological
parameters assumed are those derived from \cite{Planck2015}:
$\Omega_m~ =~ 0.307,~ \Omega_\Lambda~ =~ 0.693,~ \Omega_b ~= 0.0483,~
h = 0.678, \sigma_8 ~= ~0.823$, and $n_s ~= ~0.961$. The most striking
feature is how, for the same 7 keV sterile neutrino, the scale of the
cut-off (as measured by, say, $k_{1/4}$) changes with $L_6$.

\begin{figure}
\centering \includegraphics[trim=0.1cm 0.1cm 0.2cm
  0.2cm,clip=true,width=0.45\textwidth]{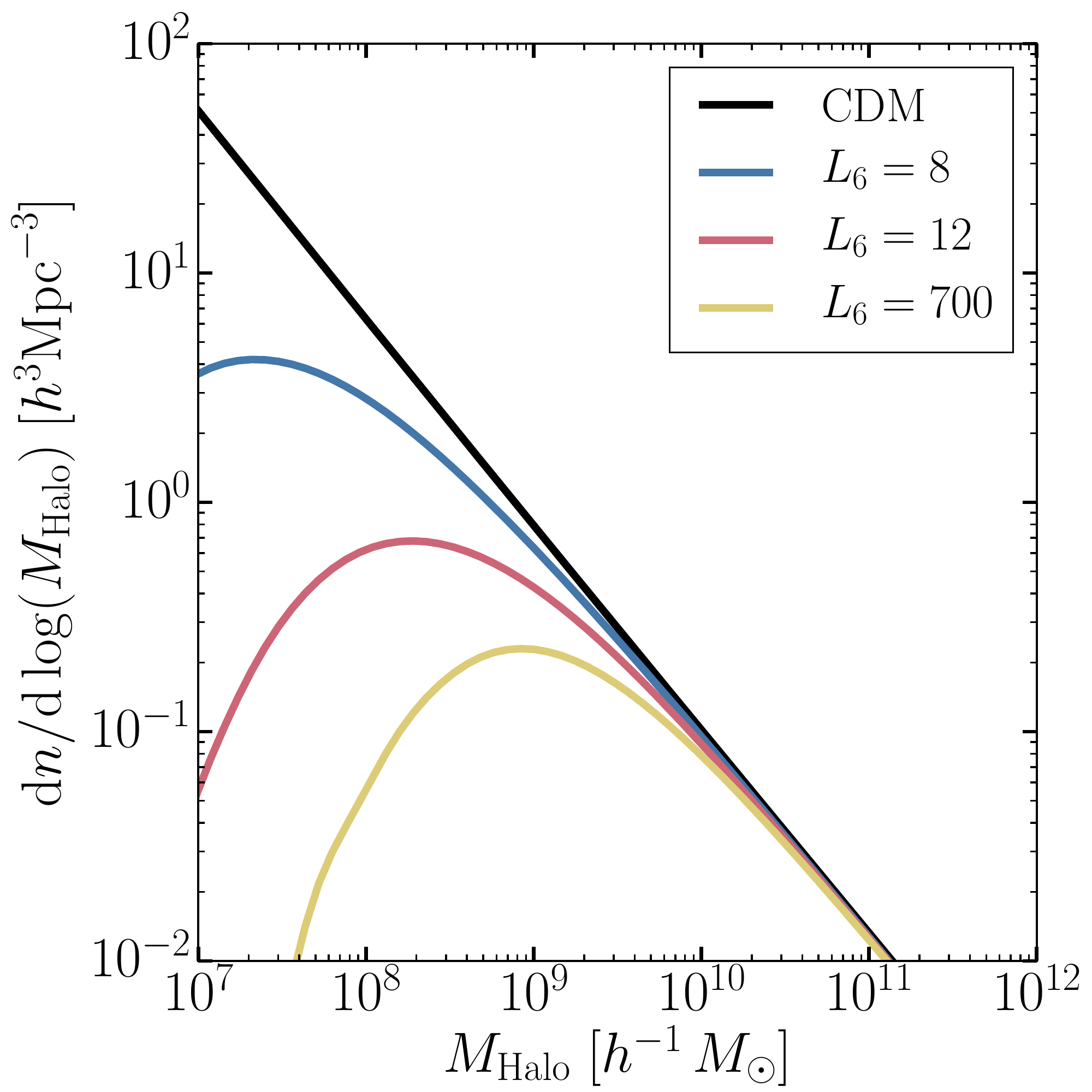}
\caption{The $z=0$ halo mass functions for CDM and 7.5~keV sterile
  neutrino models with leptogenesis parameter, $L_6 =
  \left(8,12,700\right)$, as predicted by the ellipsoidal collapse
  model of \citealt{Sheth1999}, calculated using Eqs.~\ref{st} \&
  \ref{multi}. The different cut-off scales for the sterile neutrino
  power spectra in Fig.~\ref{ps_models} are reflected in the
  different mass scales at which the corresponding halo mass functions
  are suppressed below the CDM mass function.}
\label{hmf}
\end{figure}

The truncated power spectra in the three sterile neutrino models
results in a suppression in the abundance of haloes (and by extension,
the galaxies in them) at different mass scales in the different
models. This is illustrated in Fig.~\ref{hmf} where we show the
$z=0$ halo mass functions for CDM and for $L_6 =
\left(8,12,700\right)$, as predicted by the ellipsoidal collapse
formalism of \cite{Sheth1999}. In this model, the number density of
haloes within a logarithmic interval in mass ($\rm{d}n/\rm{d}\log
M_{\rm{halo}}$) is quantified by:
\bq \label{st} \frac{\mathrm{d}n}{\mathrm{d}\log M_{\rm{halo}}} =
\frac{\bar{\rho}}{M_{\rm{halo}}}f\left(\nu\right)\left|\frac{\mathrm{d\log
    \sigma^{-1}}}{\mathrm{d}\log M_{\rm{halo}}}\right|\;, \eq
where $\bar{\rho}$ is the mean matter density of the Universe, $\nu =
\delta_c/\sigma(M_{\rm{halo}})$, $\delta_c=1.686$ is the density
threshold required for collapse and $\sigma(M_{\rm{halo}})$ is the
variance of the density field, smoothed at a scale, $M_{\rm{halo}}$
(see Section~\ref{sterilemerger}). In the ellipsoidal collapse model
the multiplicity function, $f\left(\nu\right)$, takes the form:
\bq \label{multi} f\left(\nu\right) = A \sqrt{\frac{2q\nu} {\upi}}
\left[1 + \left(q\nu\right)^{-p}\right] e^{-q\nu/2}, \eq
where $A = 0.3222, q = 0.707$ and $p = 0.3$.  Fig.~\ref{hmf} shows
how the mass functions in the sterile neutrino models peel off from
CDM at different mass scales directly related to $k_{1/4}$. The halo
masses corresponding to these wavenumbers can be estimated by:
\bq M_{1/4} = \frac{4}{3} \upi \bar{\rho}
\left(\frac{\upi}{k_{\rm{hm}}} \right)^3, \eq
giving $M_{1/4} = \left(1.1\times 10^8,7.8 \times 10^8, 2.3 \times
10^9\right)~h^{-1}~M_\odot$ for $L_6=\left(8,12,700\right)$
respectively. Clearly, the largest suppression in halo abundance
relative to CDM occurs for the $L_6 = 700$ case, and the least for the
$L_6 = 8$ case, consistent with our discussion of the significance of
the characteristic scale $k_{1/4}$. For example, at $z=0$, there are
half as many $\sim 10^8\, h^{-1}\,M_\odot$ in $L_6=8$ as in CDM. By
comparison, there are $\sim 150$ times fewer haloes at the same mass
scale for $L_6=700$ relative to CDM. The $L_6 =12$ model lies in
between these two cases, producing $\sim 20$ times fewer haloes of
$10^8\,h^{-1}\,M_\odot$.
 
\begin{table*} \centering \caption{Properties of the four dark matter
    models studied in this paper: CDM and 7 keV sterile neutrino
    models with lepton asymmetry, $L_6 =\left(8,12,700\right)$. The
    quantity $k_{1/4}$ is the wavenumber at which the amplitude of the
    power spectrum is $1/4$ that of the CDM amplitude; it is a measure
    of the ``warmth'' of the model. The last three columns indicate
    whether the model gives (1) the correct dark matter density; (2)
    whether the particle can decay to produce a line at 3.5 keV; and
    (3) whether the corresponding mixing angle can produce an X-ray
    decay flux consistent with the observations of
    \citealt{Boyarsky2014a,Bulbul2014a}.}

\begin{tabular}{ccccc}
\hline \hline\\ Model; $L_6$ & $k_{1/4}$ & Right DM abundance?  &
Decay at 3.5 keV? & Flux consistent with 3.5 keV X-ray line? \\ &
$\left[h/\rm{Mpc}\right]$ & & & \\ \hline \hline \\ CDM; $-$ & $-$ &
\cmark & \xmark & \xmark \\ 7 keV; 8 & 44.14 & \cmark & \cmark &
\cmark \\ 7 keV; 12 & 23.27 & \cmark & \cmark & \cmark \\ 7 keV; 700 &
16.05 & \cmark & \cmark & \xmark \\

\\ \hline 
\end{tabular}
\label{dmprops}
\end{table*}

\section{Galaxy formation}
\label{galaxy}

We begin by discussing the astrophysical quantities and observables
that we will use to constrain sterile neutrino models. We then briefly
introduce the semi-analytic model of galaxy formation,
\textsc{galform}, that we will use to predict these quantities for
both CDM and sterile neutrino models. We build upon the ideas and
methods laid out by \citet[][hereafter Hou15]{Hou2015}.

\subsection{A galactic ``tug-of-war''}
\label{tow}

One of the most important physical processes involved in galaxy
formation is supernova feedback (SNfb). By ejecting cold gas from
galaxies, SNfb regulates star formation, inhibiting galaxy formation
in small mass haloes \citep{Larson1974, White1991}. SNfb is thought to
be responsible for the relatively flat galaxy stellar mass and
luminosity functions compared to the steeply rising halo mass function
predicted by $N$-body simulations for $\Lambda$CDM
\citep[e.g.][]{Jenkins2001, Tinker2008, Kauffmann1993,Cole1994}. On
the smallest scales, SNfb, in conjunction with photoionisation of gas
in the early Universe, can explain the small number of faint satellite
galaxies seen around galaxies like the Milky Way in this model
\citep{Efstathiou1992, Benson2003, Sawala2015}.

Unless AGN contribute a significant number of ionising photons
\citep{Madau2015}, SNfb cannot be so strong as to suppress the
production of ionising photons at high redshift required to reionise
the Universe by $z\sim 6$, as inferred from QSO absorption lines
\citep{Mitra2015,Robertson2015} and the microwave background data
\citep{Planck2015}. Thus, at least in CDM, the small observed number
of faint galaxies sets a lower limit to the strength of feedback,
while the requirement that the Universe be ionised early enough sets
an upper limit. \cite{Hou2015} found that the simple models of SNfb
usually assumed in semi-analytic models of galaxy formation do not
satisfy both these requirements. They proposed instead a more
complicated model in which the strength of SNfb evolves in redshift,
as suggested by the SNfb model of \cite{Lagos2013} (see
Section~\ref{galjun} below).

Since in WDM the number of small haloes is naturally suppressed, for a
model to be viable, SNfb must be weak enough so that there are enough
ionising photons at high redshift, as well as a sufficient number of
satellite galaxies to account for observations.

\subsection{Supernova feedback in GALFORM}
\label{galjun}

The Durham semi-analytic model of galaxy formation, \galform~, was
introduced by \cite{Cole2000} and has been upgraded regularly as our
understanding of the physical processes involved in galaxy formation
improves and better observational constraints are obtained. For
example, \cite{Baugh2005} introduced a top-heavy IMF in bursts,
\cite{Bower2006} introduced AGN feedback and \cite{Lagos2011b}
introduced a star formation law that depends on the molecular gas
content of the ISM. The most recent version of the model
\cite{Lacey2015} includes all of these revisions.

The observational data normally used to constrain and test
semi-analytic models includes galaxies with stellar mass, $M_\star
\gsim 10^{8} {\rm M}_\odot$. When attempting to extend the
\cite{Lacey2015} model to lower mass galaxies, \cite{Hou2015} found
that the original prescription for SNfb had to be modified as
discussed in Section~\ref{tow}. In the original prescription, the mass
loading factor, $\beta$, defined as the ratio of the mass ejection
rate to the star formation rate, is assumed to be a power law in the
circular velocity, $V_{\rm{circ}}$, of the galaxy. To match the
observed satellite luminosity function and produce an acceptable
metallicity-luminosity relation for Milky Way satellites, Hou15
required a mass loading factor given by a broken power law with a
redshift dependence:
\bq \label{twopower} \beta = \begin{cases}
  \left(\rm{V}_{\rm{circ}}/\rm{V}_{\rm{SN}}\right)^{-\gamma_{\rm{SN}}}
  & \rm{V}_{\rm{circ}} \geq \rm{V}_{\rm{thresh}}
  \\ \left(\rm{V}_{\rm{circ}}/\rm{V}'_{\rm{SN}}\right)^{-\gamma'_{\rm{SN}}}
  & \rm{V}_{\rm{circ}} < \rm{V}_{\rm{thresh}},
\end{cases}
\eq
where $\rm{V}'_{\rm{SN}}$ is chosen such that the two power laws in
Eq.~\ref{twopower} join at $\rm{V}_{\rm{circ}} =
\rm{V}_{\rm{thresh}}$, $\gamma_{\rm{SN}}=3.2$, $\gamma'_{\rm{SN}} =
1.0$, $\rm{V}_{\rm{thresh}} = 50\,\rm{kms}^{-1}$ and:
\bq \label{vofz}
\rm{V}_{\rm{SN}} = \begin{cases}
180 & z > 8\\
-35z + 460  & 4 \leq z \leq 8\;.\\
320 & z < 4
\end{cases}
\eq
This redshift dependence is chosen to capture the overall behaviour of
\cite{Lagos2013} supernova feedback model. In the \cite{Hou2015}
model, the feedback strength is assumed to be the same as in
\cite{Lacey2015} at $z < 4$, but is weaker at higher redshifts and in
galaxies with $\rm{V}_{\rm{circ}} < \rm{V}_{\rm{thresh}} =
50\,\rm{kms}^{-1}$. We will refer to this feedback scheme as the
`EvoFb' (evolving feedback) model.

The values of $\gamma_{\rm{SN}}$ and $\rm{V}_{\rm{thresh}}$ in this
model were calibrated for CDM and need to be recalibrated for the
sterile neutrino models that we are considering. We find that the
values $\gamma_{\rm{SN}} = 2.6$ for $L_6=700$, $\gamma_{\rm{SN}} =
2.8$ for $L_6=\left(8,12\right)$ and $\rm{V}_{\rm{thresh}} =
30\,\rm{kms}^{-1}$ for all three values of $L_6$ provide the best-fit
to the local $b_J$ and $K$-band luminosity functions, the primary
observables used to calibrate \galform~.

   \begin{figure*} \includegraphics[trim=0.8cm 0cm 0cm
  0cm,clip=true,scale=0.45]{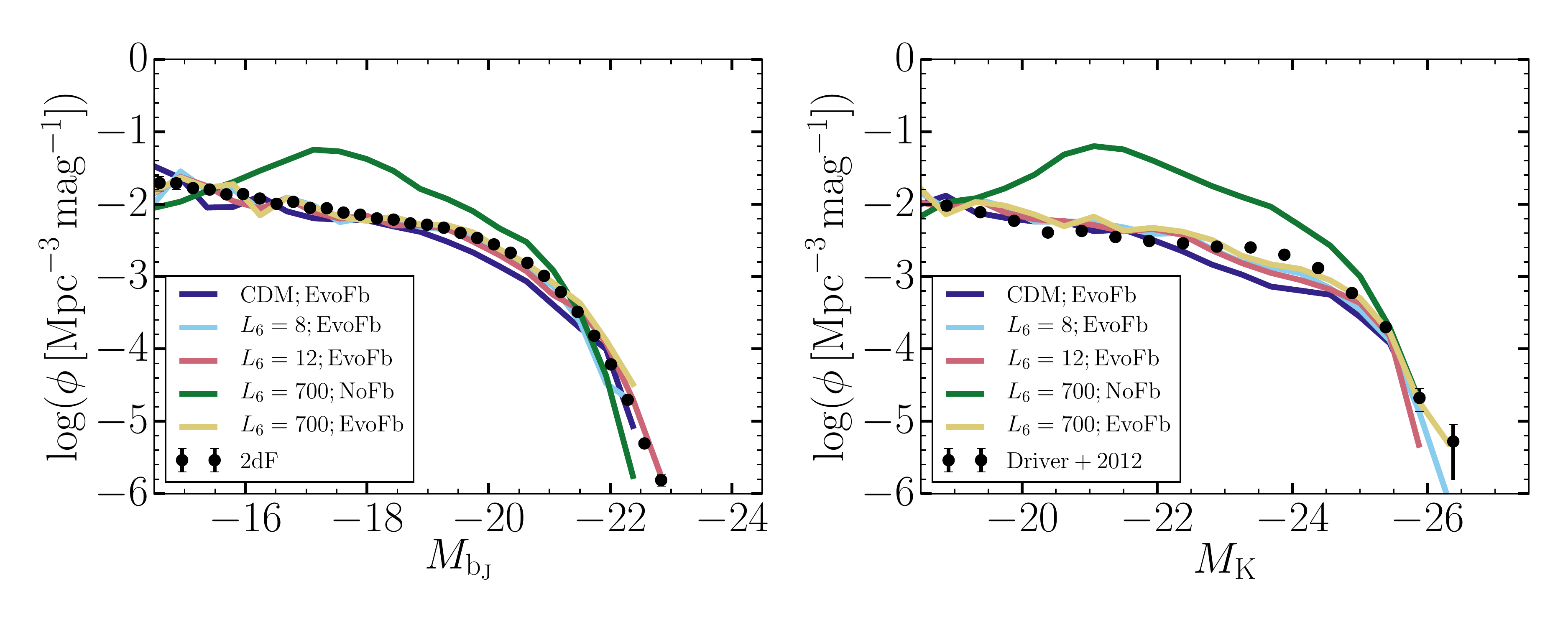} \caption{The $z=0$
       field galaxy luminosity functions in the $b_J$-band (left
       panel) and the $K$-band (right panel) for the four dark matter
       models considered in this paper: CDM and 7~keV sterile neutrino
       models with $L_6 = \left(8, 12, 700\right)$. The evolving
       feedback (EvoFb) model is used in \textsc{galform}.  For the
       $L_6 = 700$ case, we also show an extreme model in which the
       feedback has been completely turned off (`NoFb'). The black
       points are observational estimates
       \citep{Norberg2002,Driver2012}. }
  \label{fieldLF} \end{figure*}

\subsection{Halo merger trees with sterile neutrinos}
\label{sterilemerger}

We generate merger trees using the extension of the \cite{Cole2000}
Monte Carlo technique (based on the extended Press-Schechter (EPS)
theory) described in \cite{Parkinson2008}. In models in which the
linear power spectrum, $P(k)$, has a cut-off, as in our sterile
neutrino models, a small correction is required to the EPS formalism:
to obtain the variance of the density field, $\sigma(M_{\rm{halo}})$,
$P(k)$ needs to be convolved with a sharp $k$-space filter rather than
with the real-space top-hat filter used for CDM
\citep{Benson2013}. This choice results in good agreement with the
conditional halo mass function obtained in $N$-body simulations
\cite[see, for example, Fig. 6 in][]{Lovell2015}.

Using our Monte Carlo technique rather than $N$-body simulations to
generate merger trees has the advantage that different sterile
neutrino models can be studied at minimum computational expense while
avoiding the complication of spurious fragmentation in filaments that
occurs in $N$-body simulations with a resolved cut-off in $P(k)$
\citep[e.g.][]{Wang2007,Lovell2014}.

\section{Results}
\label{results}

In this section, we present the main results of our models, consisting
of predictions for field and satellite luminosity functions and the
redshift of reionisation. We also investigate the sources that produce
the ionising photons at high redshift.

\subsection{Field luminosity functions}
\label{field_lfs}

As discussed in Section~\ref{galjun}, the parameters of the SNfb model
in \galform~ were calibrated so as to obtain a good match to the
present-day field galaxy luminosity functions. The $b_J$ and $K$-band
luminosity function in CDM and the $L_6 = \left(8,12,700\right)$ 7 keV
sterile neutrino models are shown in Fig.~\ref{fieldLF}. In both
cases we have made use of the EvoFb feedback scheme of
Section~\ref{galjun}. We also consider an extreme model for $L_6 =
700$, in which supernova feedback is turned off completely (`NoFb'),
thus maximising the amount of gas that is converted into stars.

In Fig.~\ref{fieldLF} we see that with the EvoFb scheme the observed
luminosity functions are well reproduced in CDM and all our sterile
neutrino models. This should come as no surprise since the EvoFb model
parameters were tuned to match these particular data. As mentioned in
Section~\ref{sterile}, the $L_6 = 700$ model, while inconsistent with
the 3.5 keV line (see Table~\ref{dmprops}), is interesting because it
has the most extreme power spectrum cut-off for a 7~keV sterile
neutrino that produces the correct dark matter abundance. The maximum
star formation efficiency in any model is obtained by turning off SNfb
altogether. If in this limiting scenario the $L_6=700$ model produces
too few faint galaxies to match the field luminosity function, this
extreme model would be strongly ruled out. As Fig.~\ref{fieldLF}
shows, the resultant luminosity function (shown in green) in fact
overproduces faint galaxies.

\begin{figure*}
\includegraphics[scale=0.35]{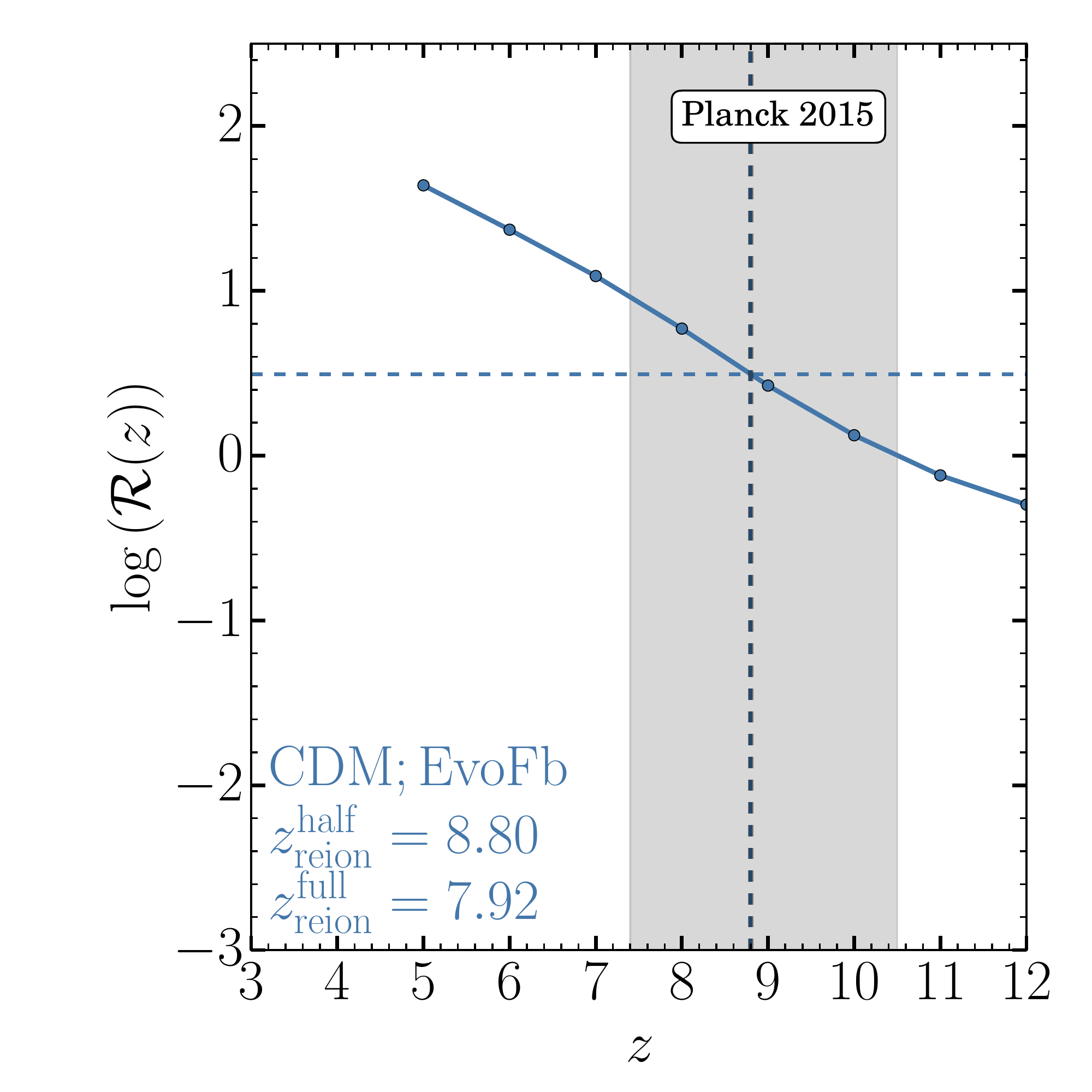}
\includegraphics[scale=0.35]{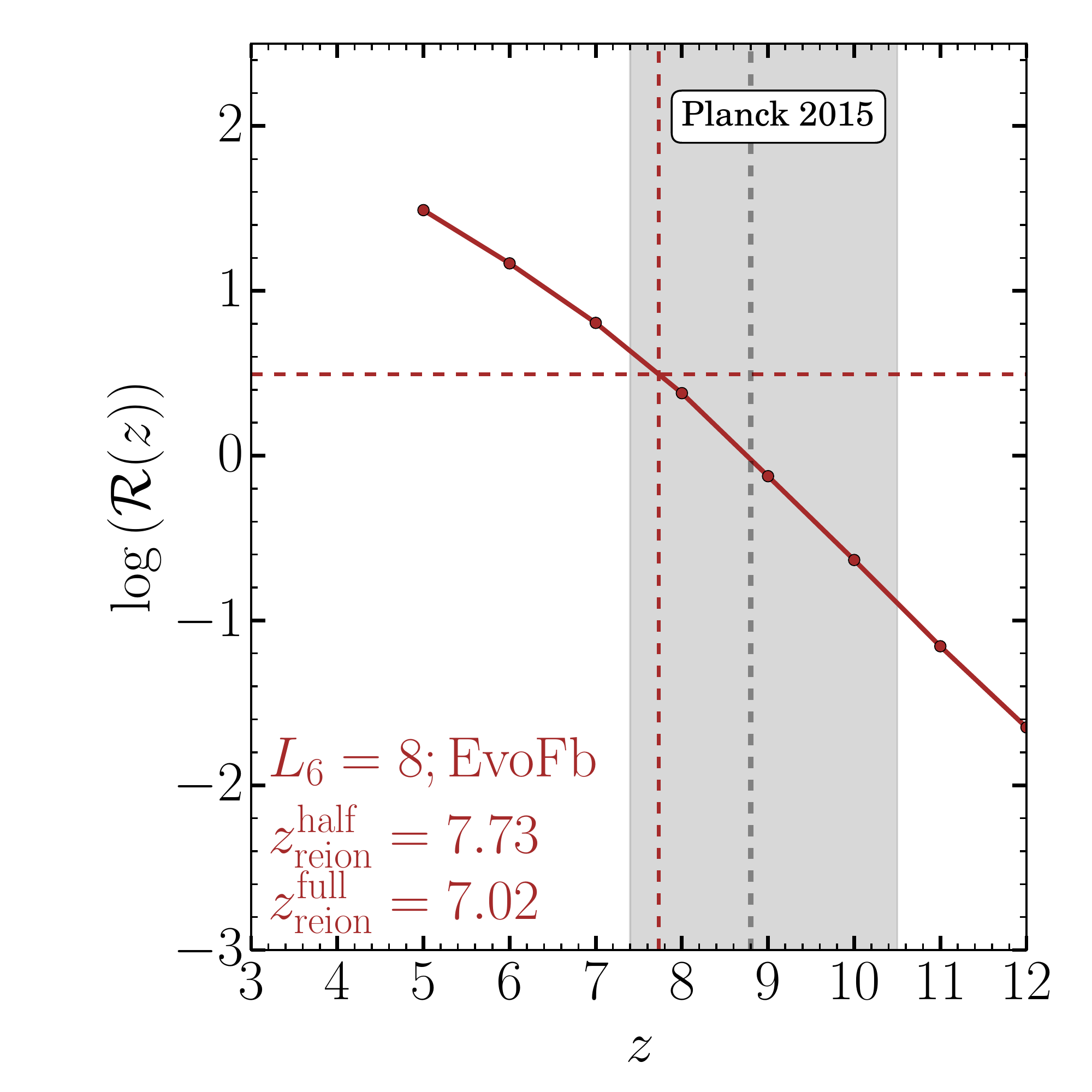}
\includegraphics[scale=0.35]{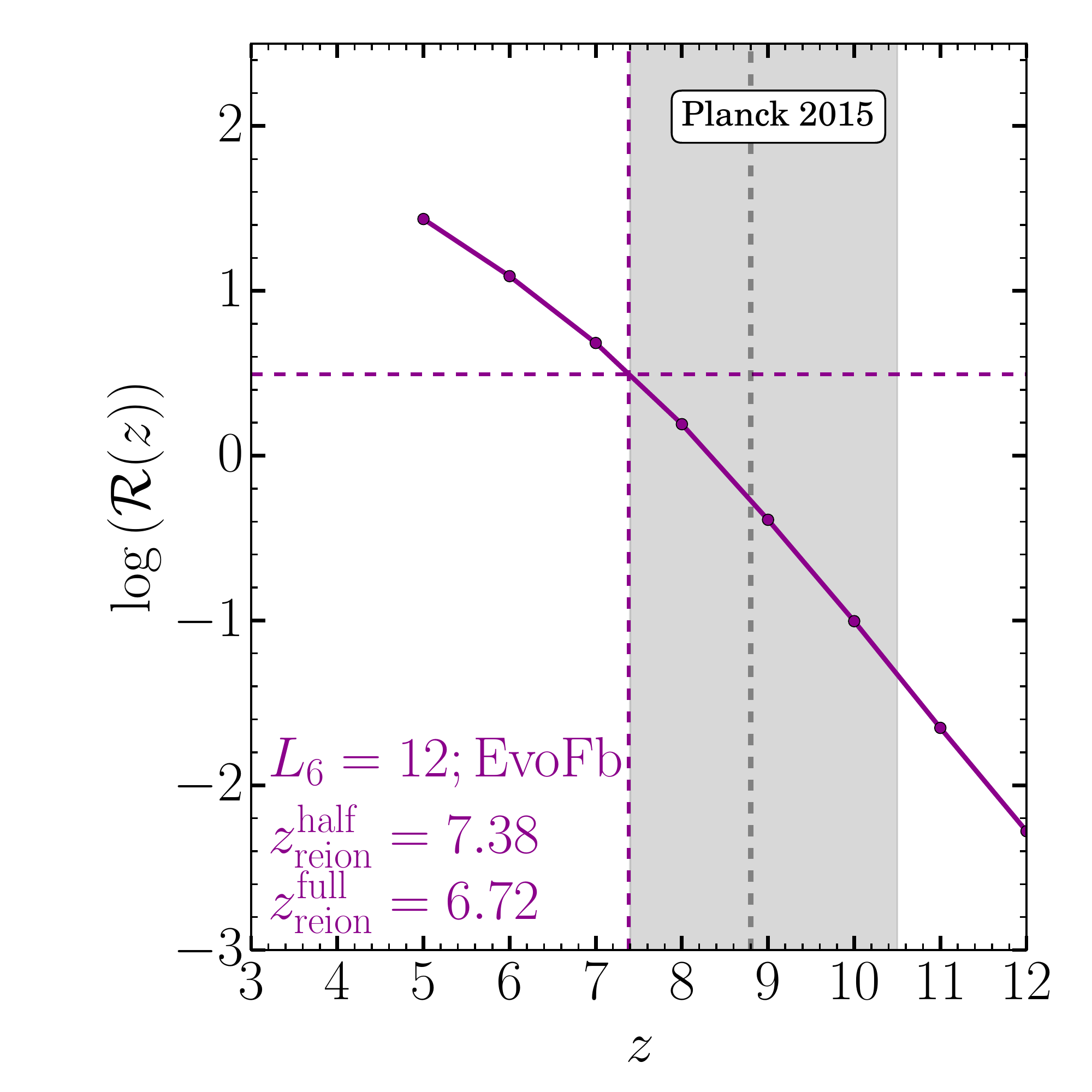}
\includegraphics[scale=0.35]{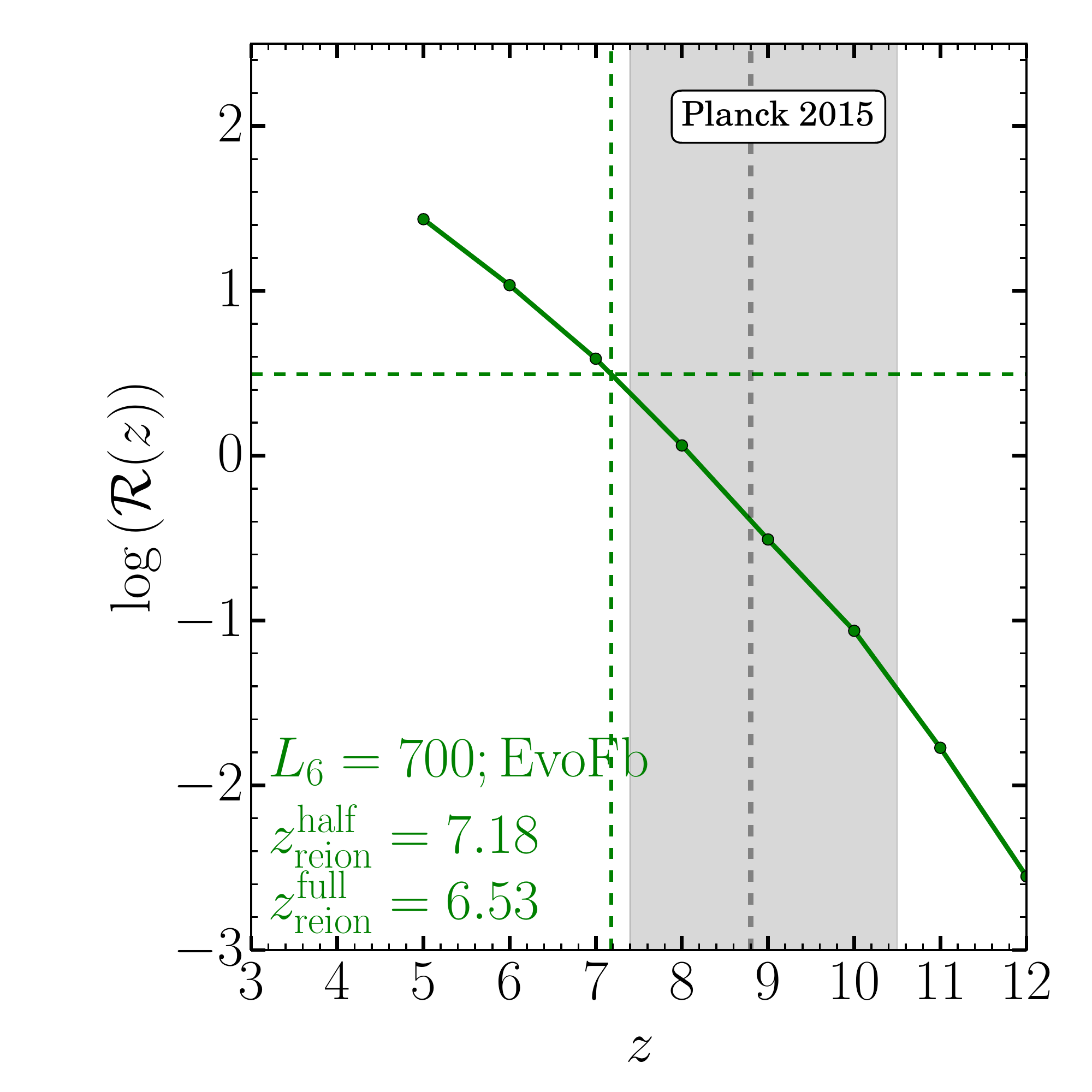}
\caption{The ratio of the total number of ionising photons produced up
  to redshift $z$ as a fraction of the total comoving number density
  of hydrogen nuclei (solid lines in each panel). In each panel, we
  show the predictions for the different dark matter models under the
  EvoFb scheme. The intersection of the coloured dashed lines marks
  the redshift at which the Universe is 50\% ionised; the redshifts
  for 50\% ($z_{\rm{reion}}^{\rm{half}}$) and 100\% reionisation
  ($z_{\rm{reion}}^{\rm{full}}$) are listed in the bottom left of each
  panel. The dashed grey line and shaded grey region demarcate the
  observational constraints as obtained from the {\it Planck}
  satellite, $z_{\rm{reion}}^{\rm{half}} = 8.8^{+1.7}_{-1.4}$ (at 68\%
  confidence).}
\label{reion}
\end{figure*}

\begin{figure*}
\includegraphics[trim=1cm 0cm 0cm
  0cm,clip=true,scale=0.48]{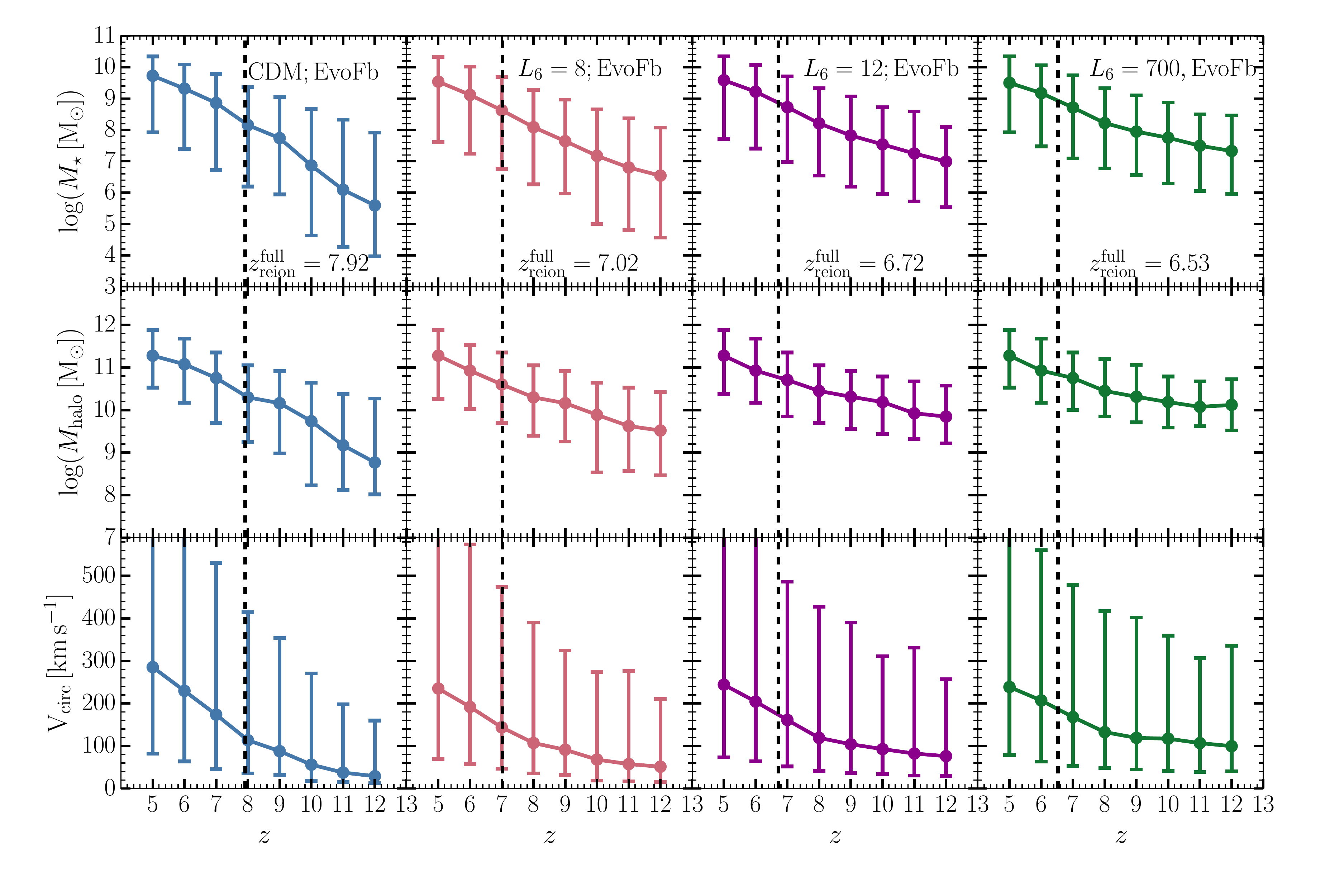}
\caption{Properties of the sources that produce ionising photons as a
  function of redshift for CDM and 7 keV sterile neutrino models with
  $L_6 = \left(8,12,700\right)$. The properties shown are stellar
  mass, $M_\star$ (top row), halo mass $M_{\rm{halo}}$ (middle row)
  and circular velocity ($\rm{V}_{\rm{circ}}$). The median (solid
  lines), 5th and 95th percentiles (error bars) are determined by
  weighting the contribution of each galaxy to the total ionising
  emissivity at that redshift. The black vertical dashed line in each
  case marks the redshift at which the Universe is half ionised. }
\label{sources}
\end{figure*}

\begin{figure*}
\includegraphics[trim=0.2cm 0cm 0cm
  0cm,clip=true,scale=0.48]{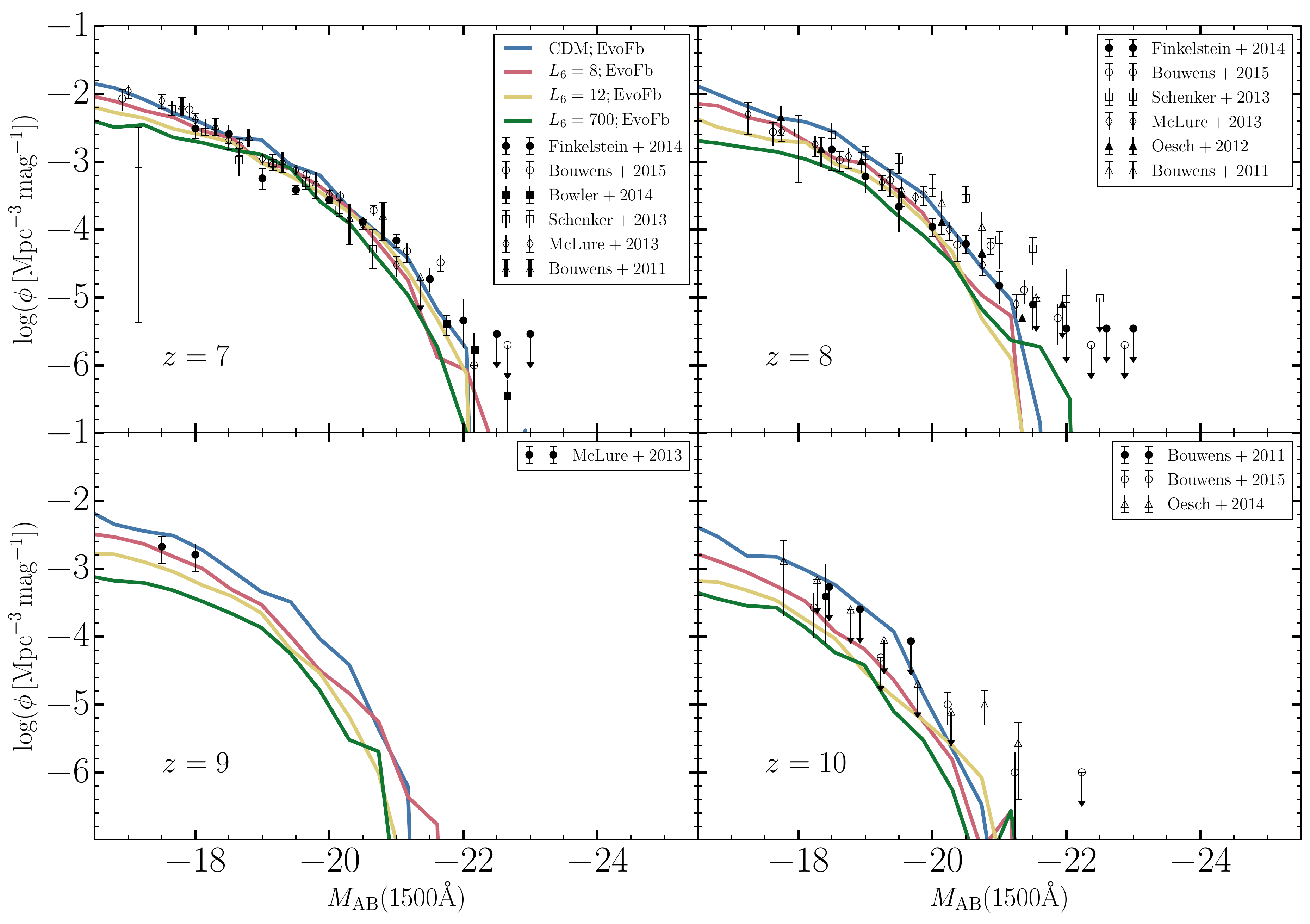}
\caption{Evolution of the rest frame far-UV galaxy luminosity
  functions from $z=7-10$ in our models. The predictions of \galform~
  for CDM and the $L_6 = \left(8,12,700\right)$ 7~keV sterile neutrino
  models are shown with solid colour lines as indicated in the legend.
  The symbols with errorbars are observational measurements
  \citep{uv_lf_bouwens2011, uv_lf_bouwens2011n, uv_lf_oesch2012,
    uv_lf_schenker, uv_lf_mclure, uv_lf_finkelstein, uv_lf_bowler,
    uv_lf_oesch2014, uv_lf_bouwens2015}.}
\label{uvlfz}
\end{figure*}

\subsection{Redshift of reionisation}
\label{z_reion}

Since the onset of halo formation occurs later in sterile neutrino
models compared to CDM \citep[e.g.][]{Bose2016}, star formation in
dwarf galaxies is delayed \citep{Governato2015}. Since, in addition,
there are no haloes below a cut-off mass, it is unclear that enough
sources of ionising photons will have formed to ionise hydrogen early
enough to be consistent with the {\it Planck} limits on the redshift
of reionisation \citep{Planck2015}.

To answer this question we use \galform~ to calculate the ratio of the
comoving number density of ionising photons produced, $n_\gamma$, to
that of hydrogen nuclei, $n_{\rm{H}}$ as:
\bq \label{ratio_photon} \mathcal{R}(z) = \frac{n_\gamma}{n_{\rm{H}}}
= \frac{\int_z^\infty \epsilon(z')\,{\rm d}z'}{n_{{\rm{H}}}}, \eq
where $\epsilon(z')$ is the comoving number density of Lyman continuum
photons produced per unit redshift. The Universe is deemed to be fully
ionised at redshift $z_{\rm{reion}}^{\rm{full}}$ when the ratio in
Eq.~\ref{ratio_photon} reaches the value:
\bq \label{photon_thresh} \left.\mathcal{R}(z)\right|_{\rm{full}} =
\frac{1+N_{\rm{rec}}}{f_{\rm{esc}}} = 6.25.  \eq
Here $N_{\rm{rec}}$ is the number of recombinations per hydrogen atom
and $f_{\rm{esc}}$ is the fraction of ionising photons that are able
to escape a galaxy into the IGM. \cite{Raicevic2011} advocate a value
of $N_{\rm{rec}} = 1$ based on the hydrodynamical simulations of
\cite{Iliev2006} and \cite{Trac2007}.  \cite{Finlator2012} suggest
that photoheating would smooth the diffuse IGM and reduce the clumping
factor by a factor of three compared with the value derived by
\cite{Iliev2006}. In this work, we will adopt a value $N_{\rm{rec}}=
0.25$ (as in Hou15), but we have checked that our conclusions are
insensitive to the exact value of this parameter. Furthermore, we
assume $f_{\rm{esc}}=0.2$, which is consistent with the value used by
\cite{Raicevic2011}.  \cite{Sharma2016} present observational and
theoretical evidence in support of this choice of $f_{\rm{esc}}$.

The microwave background data measure the optical depth to the time
when the Universe (re)combined. This is usually converted into an
equivalent `redshift of reionisation' assuming a model of
non-instantaneous reionisation. The value quoted in \cite{Planck2015}
corresponds to $z_{\rm{reion}}^{\rm{half}}$, the redshift at which the
Universe is {\it half} ionised. With our assumptions this corresponds
to:
\bq
\left.\mathcal{R}(z)\right|_{\rm{half}} = 3.125.
\eq
Reionisation suppresses galaxy formation in low mass haloes through an
effect known as photoionisation feedback. In \galform~, this is
modelled using the approximation described in \cite{Benson2003}: for
haloes with virial velocity $V_{\rm{vir}} < V_{\rm{crit}}$, no gas
cooling takes place for $z < z_{\rm{crit}}$. As in Hou15, we adopt
$z_{\rm{crit}} = z_{\rm{reion}}^{\rm{full}}$ and $V_{\rm{crit}} =
30\,\rm{kms}^{-1}$ \citep{Okamoto2008}.

In the standard \cite{Lacey2015} prescription, SNfb is modelled as a
power law in the circular velocity of the galaxy without any
dependence on redshift. Hou15 found that this model predicts
$z_{\rm{reion}}^{\rm{half}} = 6.1$ for CDM, in conflict with the
bounds by \cite{Planck2015}: $z_{\rm{reion}}^{\rm{half}} =
8.8^{+1.7}_{-1.4}$. We expect that sterile neutrino models, in which
the formation of galaxies is both suppressed and delayed, would be in
even greater conflict with the {\it Planck} observations. For this
reason, in what follows we only consider the predictions of the
evolving feedback (EvoFb) model of Hou15 (Section~\ref{galjun}) which,
at least for CDM, predicts an acceptable value for
$z_{\rm{reion}}^{\rm{half}}$.

Fig.~\ref{reion} shows the evolution of $\mathcal{R}(z)$ with
redshift for CDM and sterile neutrino models with $L_6 =
\left(8,12,700\right)$ according to \galform~ with EvoFb feedback. In
each panel, the intersection of the colour dashed lines marks
$z_{\rm{reion}}^{\rm{half}}$, where $n_\gamma/n_{\rm{H}} = 3.125$. The
dashed grey line and shaded grey region mark the median and 68\%
confidence intervals from \cite{Planck2015}:
$z_{\rm{reion}}^{\rm{half}} = 8.8^{+1.7}_{-1.4}$. In the bottom left
of each panel, we give $z_{\rm{reion}}^{\rm{half}}$ and
$z_{\rm{reion}}^{\rm{full}}$ predicted for each model.

All three 7~keV sterile neutrino models have values of
$z_{\rm{reion}}^{\rm{half}}$ that are broadly consistent with the {\it
  Planck} data. The $L_6 = \left(12, 700\right)$ models fall just
outside the lower 68\% confidence lower limit and the $L_6=8$ model
just inside. This is a non-trivial result given the paucity of early
structure in these models compared to CDM. Unsurprisingly,
$z_{\rm{reion}}^{\rm{half}}$ is higher in CDM \footnote{We note that
  our results in this section contradict those by
  \cite{Rudakovskiy2016}, who find that in the 7 keV $L_6=10$ model
  the Universe is reionised {\it earlier} than in CDM.  This is
  ascribed to the lack of `mini'-haloes in the sterile neutrino
  cosmology, which reduces the average number of recombinations per
  hydrogen atom. In our analysis this amounts to a reduction in the
  value of $N_{\rm{rec}}$ in Eq.~\ref{photon_thresh}. However, we have
  checked that even reducing the value of $N_{\rm{rec}}$ by a factor
  of 10 does not affect our results significantly.}.  Fig.~\ref{reion}
already hints at the reason why the sterile neutrino models are able
to ionise the Universe early enough. Comparing, for example, the
$L_6=700$ model (bottom right panel) to CDM (top left panel), it is
clear that the evolution of $\log\left(\mathcal{R}(z)\right)$ is
steeper in the former, that is more UV photons are produced per unit
redshift in the $L_6=700$ case, even though the {\it total} number of
photons at that redshift is larger in CDM. For $L_6=8$, the most
`CDM-like' sterile neutrino model, the gradient of
$\log\left(\mathcal{R}(z)\right)$ is shallower. We will return to this
feature shortly.

\subsection{The galaxies responsible for reionisation}
\label{sources_reion}

We have seen that in spite of the delayed onset of galaxy formation,
even the most extreme 7 keV sterile neutrino model is able to ionise
the Universe early enough to be consistent with the constraints from
{\it Planck}. To explore why this is so, we show in Fig.~\ref{sources}
several properties of the sources that contribute the bulk of the
ionising photons at each redshift. Each column in the figure
corresponds to a different dark matter model, while each row
corresponds to a different property of the ionising sources: total
stellar mass ($M_\star$, first row), halo mass ($M_{\rm{halo}}$,
second row) and galaxy circular velocity ($\rm{V}_{\rm{circ}}$, third
row). The black vertical dashed lines mark
$z_{\rm{reion}}^{\rm{full}}$, which is given in the top row in each
case.

In CDM, the median stellar mass (i.e. the mass below which galaxies
produce 50\% of the ionising emissivity) at
$z=z_{\rm{reion}}^{\rm{full}}$ is $\sim 10^8\,M_\odot$, whereas in the
three sterile neutrino models the median mass is close to $\sim
10^9\,M_\odot$. The larger scatter in $M_\star$ and $M_{\rm{halo}}$
for CDM is due to the wide range of mass of the galaxies that
contribute to the ionising photon budget. For example, at $z=10$,
galaxies with mass in the range $10^4\,M_\odot < M_\star <
10^9\,M_\odot$ contribute 90\% of the ionising photons, whereas in the
$L_6=\left(12,700\right)$ models, 90\% of the photons are produced by
galaxies with mass in the range $10^6\,M_\odot < M_\star <
10^9\,M_\odot$ since very few galaxies with $M_\star < 10^6\,M_\odot$
form in these models. The result is that the primary sources of
ionising photons at high redshift in sterile neutrino are on average
{\it more massive} than in CDM.

The build-up of the galaxy population in our models is illustrated in
Fig.~\ref{uvlfz} which shows the rest frame far-UV (1500 \AA)
luminosity functions at $z=7,8,9,10$ in CDM and the
$L_6=\left(8,12,700\right)$ models. As noted in Hou15, in CDM the
EvoFb feedback model predicts luminosity functions that are in good
agreement with the data at all redshifts. The same is true for $L_6=8$
but the $L_6=\left(12,700\right)$ models underpredict the abundance of
galaxies fainter than $M_{\rm{AB}} (1500\rm{\AA}) \sim -20$ galaxies
at $z=9$ and $z=10$. Reducing the strength of SNfb at $z>8$ slightly
can bring these models into agreement with the data without spoiling
the agreement at $z=0$.

An interesting feature of Fig.~\ref{uvlfz} is that while the
$L_6=\left(8,12,700\right)$ sterile neutrino models produce fewer
galaxies fainter than $M_{\rm{AB}} (1500\rm{\AA}) \sim -20$ at $z=10$,
all three models catch up with CDM by $z=7$, roughly the time by which
50\% hydrogen reionisation has occurred. The {\it build-up of the high
  redshift galaxies therefore proceeds more rapidly} in the sterile
neutrino cosmologies than in CDM. This is consistent with the
behaviour of the rate of ionising photon production seen in
Section~\ref{z_reion}, where the slope of
$\log\left(n_\gamma/n_{\rm{H}}\right)$ was shown to be steeper for
sterile neutrino models compared to CDM.

The reason for the differing rates of galaxy formation at high
redshift in the different models can be understood as follows. Due to
the lack of progenitors below the cut-off mass scale, WDM haloes build
up via roughly equal-mass mergers of intermediate mass haloes. Near
the free streaming scale, the growth rate of haloes is therefore more
rapid in WDM than in CDM \cite[see, e.g.][]{Ludlow2016}. This is why
soon after the formation of the first galaxies the rate of galaxy
formation in sterile neutrino models `catches up' with the
corresponding rate in CDM. This rapid early evolution, reflected for
example in the UV luminosity function, is a generic prediction of WDM,
independently of the details of the galaxy formation model.

\begin{figure*}
\includegraphics[scale=0.4]{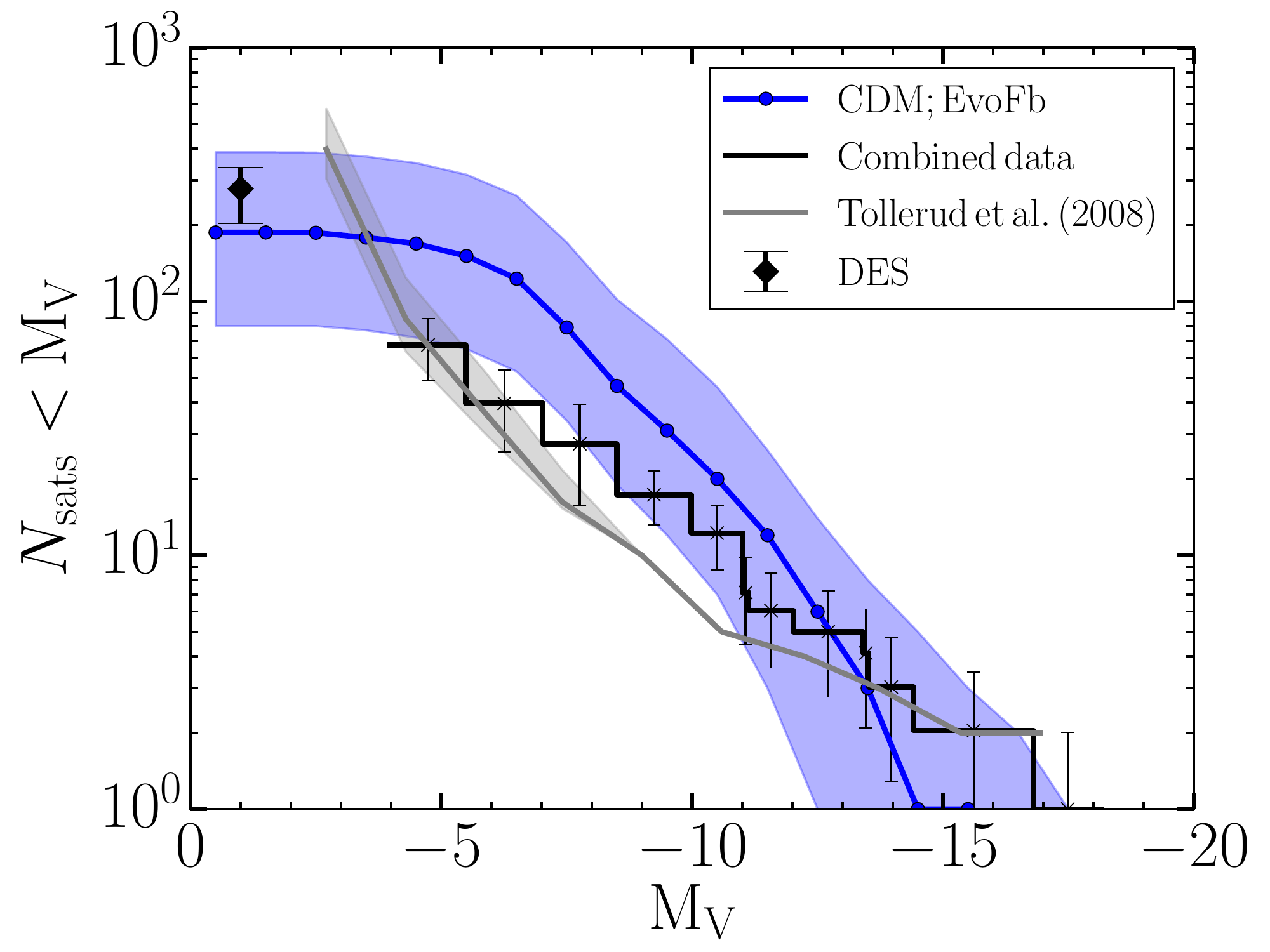}
\includegraphics[scale=0.4]{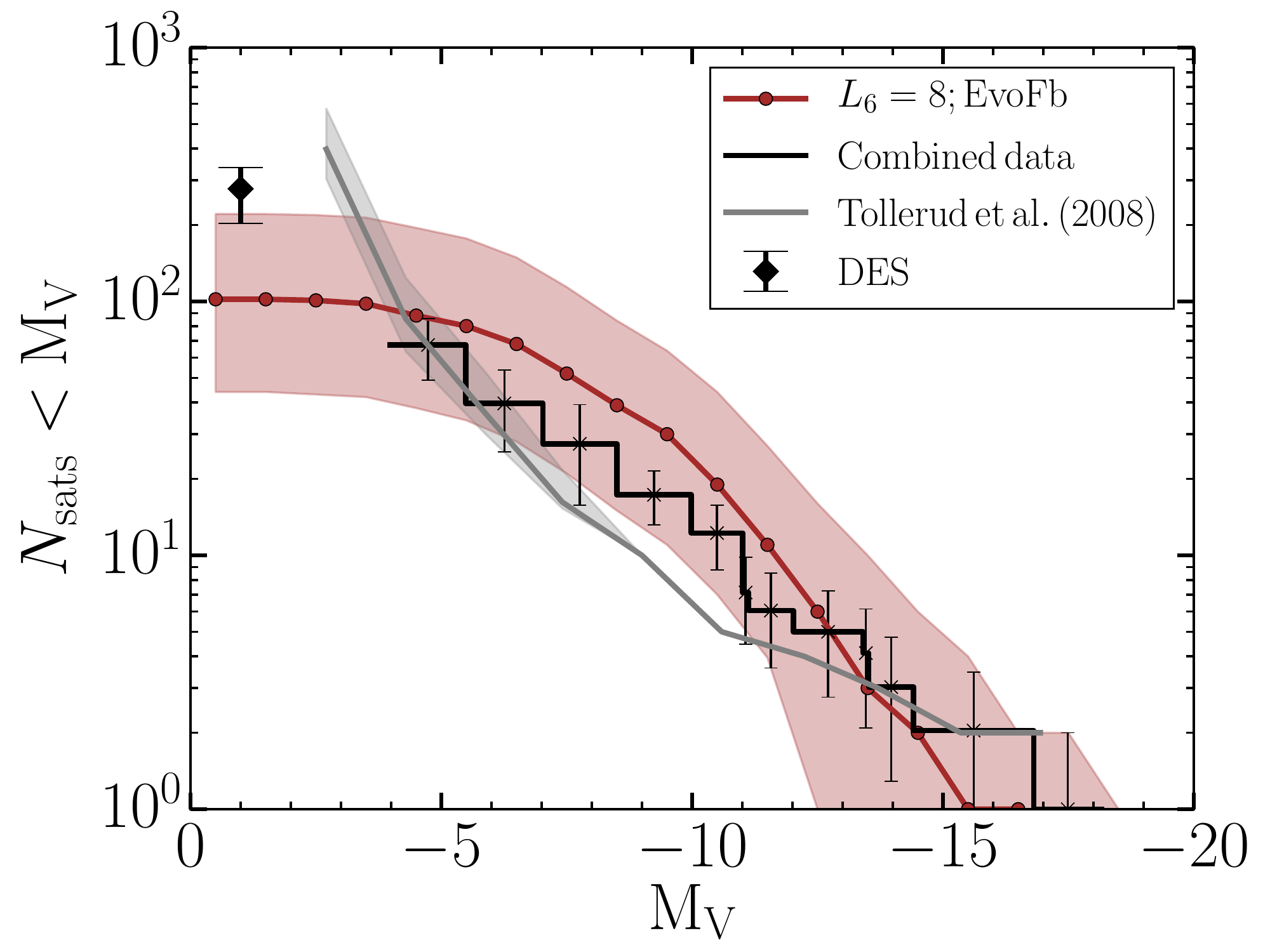}
\includegraphics[scale=0.4]{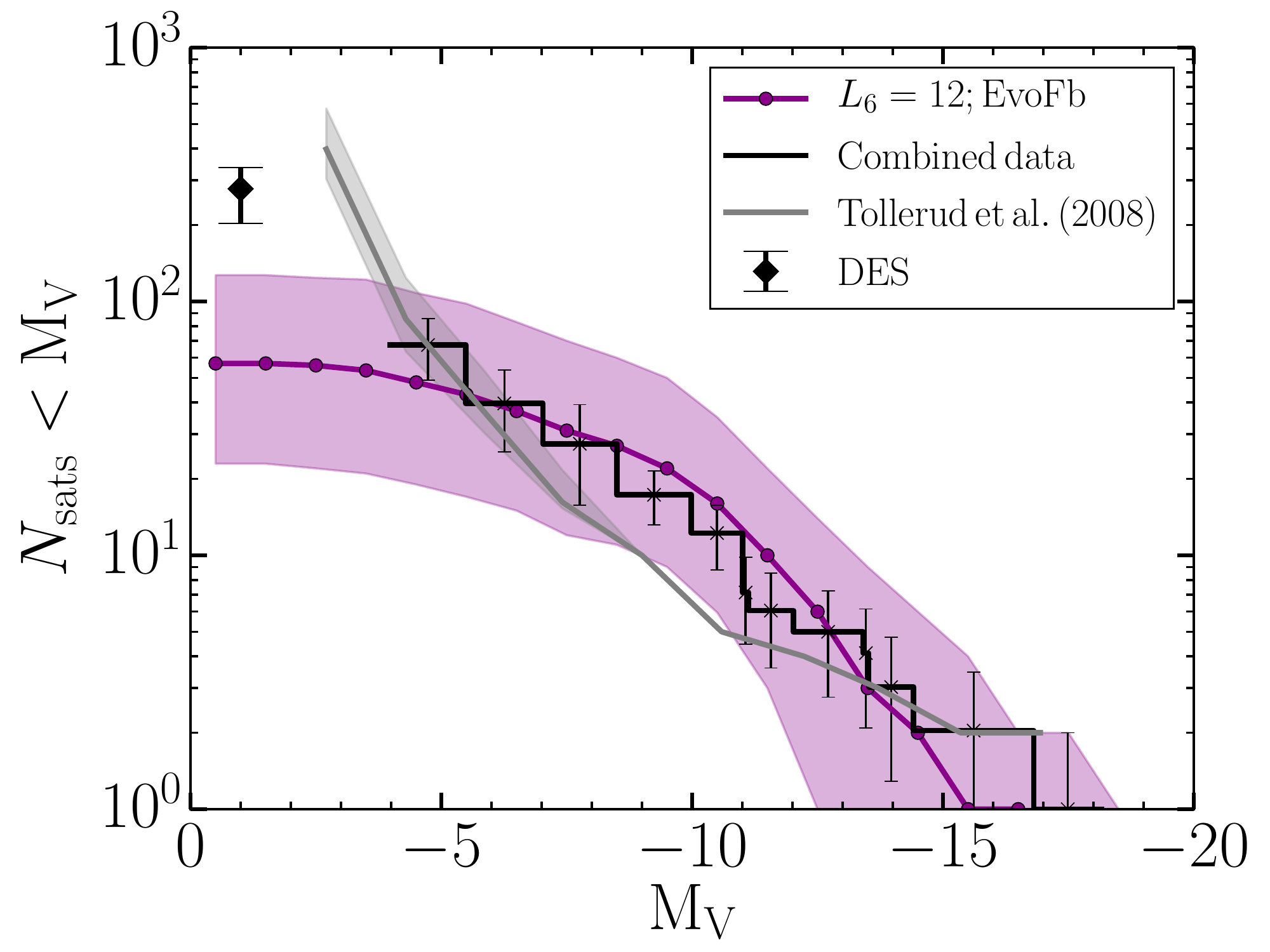}
\includegraphics[scale=0.4]{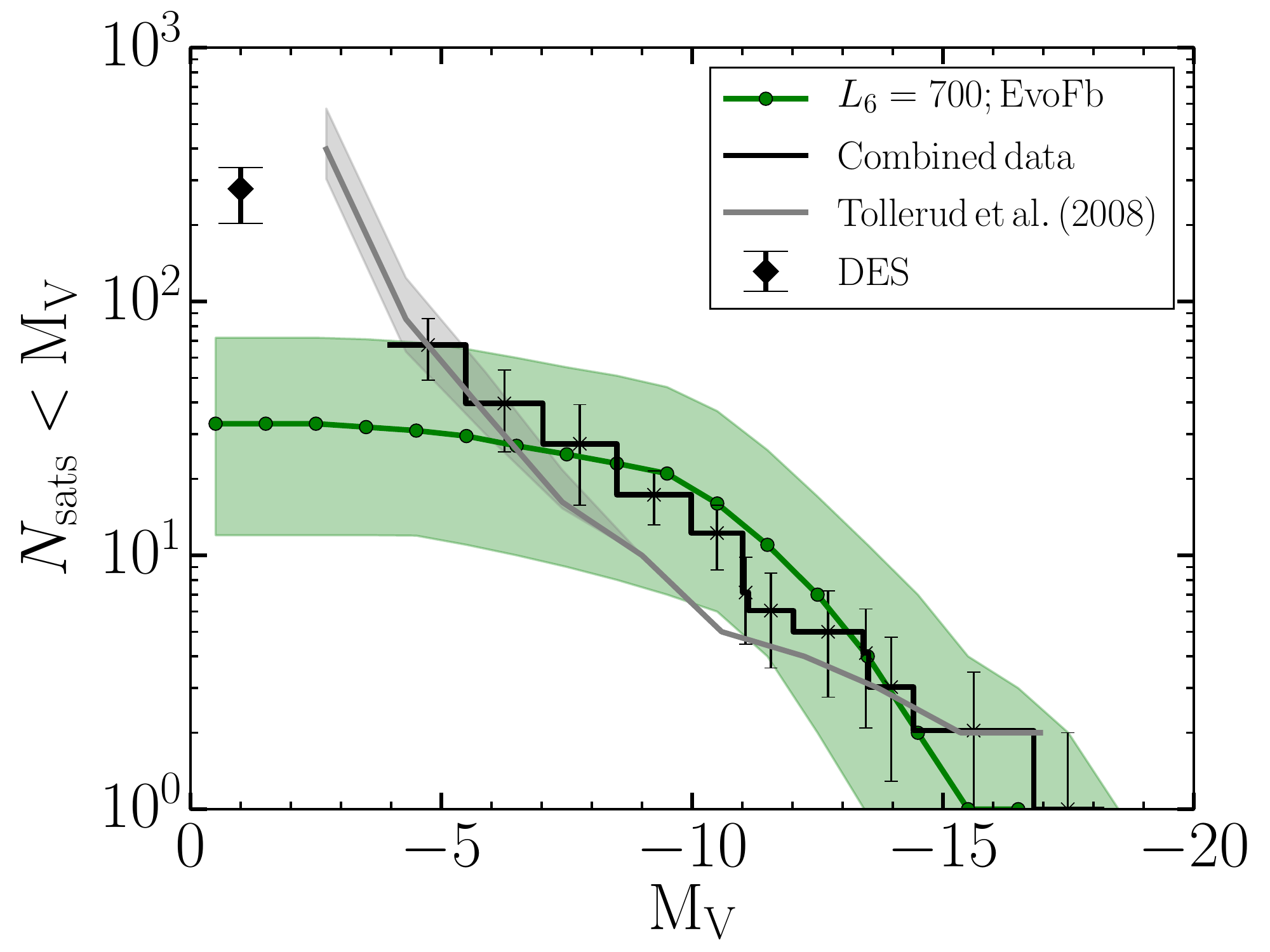}
\caption{Cumulative $V$-band Milky Way satellite luminosity functions
  at $z = 0$ for our four dark matter models with EvoFb supernova
  feedback. In each case, we have used 100 Monte Carlo merger trees
  for haloes of final mass in the range $5 \times 10^{11} - 2 \times
  10^{12}\,M_\odot$. The smooth solid line indicates the median and
  the coloured shaded region the 5th and 95th percentiles over all
  realisations. The black histogram labelled `Combined data' shows the
  observed Milky Way satellite luminosity function obtained by
  combining two datasets: for $M_{\rm{V}} \geq -11$ the data are taken
  from \citet{Koposov2008}, which includes corrections for
  incompleteness in the SDSS DR5 catalogue; for $M_{\rm{V}} < -11$,
  the data are taken from \citet{McConnachie2012}. The solid grey line
  shows the satellite luminosity function from \citet{Tollerud2008}
  with the grey shaded region showing the 98\% spread over 18,576 mock
  surveys of the Milky Way halo in the Via Lactea simulation
  (\citealt{Diemand2007}).  The black diamond marks an extension of
  the observed satellite luminosity function adding the new
  ultra-faint dwarf satellites discovered by DES down to $M_{\rm{V}}
  \leq -1$ (\citealt{Jethwa2016}). The partial sky coverage of the
  survey is taken into account.  All error bars are Poisson errors,
  including volume corrections where appropriate.}
\label{MWsats}
\end{figure*}

\subsection{Satellites of the Milky Way}
\label{sats_of_mw}

The Milky Way satellite luminosity function has been used to set
limits on the warm dark matter particle mass: if the power spectrum
cut-off occurs on too large a scale, too few haloes form to account
for the observed number of satellites
\citep{Maccio2010,Polisensky2011,Lovell2012,Nierenberg2013,Kennedy2014}.
These studies considered non-resonantly produced thermal relics
\citep[but see][]{Schneider2016}.  \cite{Lovell2015} considered
sterile neutrino models, similar to ours, with different particle
masses and values of $L_6$ and an earlier version of \galform~
\citep{GonzalezPerez2014}. There are degeneracies between the shape of
the WDM power spectrum and some of the parameters of the galaxy
formation model, particularly, of course, the strength of SNfb (see
\citealt{Kennedy2014} for a discussion). These degeneracies are
mitigated in our case by considering a variety of observational
constraints involving a range of halo masses and redshifts.

We have allowed the strength of SNfb to vary with redshift, by
assuming that SNfb is weaker at high redshift. In
Section~\ref{z_reion}, we found that this modification to the feedback
scheme in \galform~ allows CDM and the $L_6 = \left(8,12,700\right)$
sterile neutrino models to reionise the Universe early enough to be
consistent with the {\it Planck} limits on the redshift of
reionisation. It is not clear, however, what the effect of reducing
the strength of feedback will be on observables at lower redshifts. In
particular, we expect the predicted luminosity function of satellites
in the Milky Way to be particularly sensitive to this modification.

To predict the satellite luminosity functions around galaxies similar
to the Milky Way we generate 100 Monte Carlo merger trees in 5 equally
spaced bins of final halo masses in the range $5 \times 10^{11}\,
M_\odot \leq M_{\rm{halo}}^{\rm{host}} \leq 2 \times 10^{12}\,
M_\odot$. The cumulative $V$-band satellite luminosity functions at
$z=0$ are shown in Fig.~\ref{MWsats} for our various dark matter
models with the EvoFb feedback scheme. Before we attempt to compare
these predictions with observations we note that the two different
observational datasets plotted in the figure disagree with one another
at the bright end of the luminosity function ($M_{\rm{V}} \leq -8$),
which is the regime of the 11 ``classical'' satellites. There are two
reasons for this difference: firstly, \cite{McConnachie2012}, whose
measurements are included in the bright end of the `Combined data'
sample includes Canis Major ($M_{\rm{V}} = -14.4$), whereas this
galaxy is excluded by \cite{Tollerud2008}. Secondly,
\cite{Tollerud2008} adopt $M_{\rm{V}} = -9.8$ for Sculptor, compared
to McConnachie's value of $M_{\rm{V}} = -11.1$. At the faint end the
differences in the satellite luminosity function arise from differing
assumptions for the radial distributions of the satellites. In
particular, \cite{Koposov2008} assume that the satellite distribution
follows the NFW profile \citep{Navarro1996,Navarro1997} of the host
halo, whereas \cite{Tollerud2008} assume the subhalo radial
distribution measured in the Via Lactea simulations
\citep{Diemand2007}. The radial distribution of subhaloes is similar
in CDM and WDM \citep{Bose2016b}.

Fig.~\ref{MWsats} shows that all of our models, including the most
extreme $L_6=700$ case, are consistent with the data down to
$M_{\rm{V}} \sim -5$. For CDM the EvoFb model slightly overpredicts
the number of the faintest satellites ($M_{\rm{V}}>-8$), but here the
data could be incomplete.  However, since the number of satellites
scales with the host halo mass \citep{Wang2012,Cautun2014}, our
sterile neutrino models would be increasingly in conflict with the
observed luminosity functions for $M_{\rm{halo}}^{\rm{host}} \leq
10^{12}\, M_\odot$. For example, if $M_{\rm{halo}}^{\rm{host}} \leq 7
\times 10^{11}\, M_\odot$, both the $L_6 = 700$ and $L_6 = 12$ EvoFb
models would be ruled out because they fail to form enough faint
satellites with $M_{\rm{V}} > -10$ even after accounting for the large
scatter. Only CDM and our $L_6=8$ sterile neutrino models would remain
consistent with the \cite{Koposov2008} and \cite{McConnachie2012}
(`Combined data') observations in this case.

The Dark Energy Survey (DES) recently reported the discovery of new
ultra-faint dwarf galaxies
\citep{Bechtol2015,Koposov2015,DrlicaWagner2015, Jethwa2016}.  We can
consider their contribution to the observed luminosity function
following the analysis by \cite{Jethwa2016} who find that 12 of the 14
satellites have $> 50\%$ probability of having been brought in as
satellites of the LMC itself (at 95\% confidence).  Extrapolating from
the detected population \cite{Jethwa2016} conclude that the Milky Way
should have $\sim 180$ satellites within $300\,\rm{kpc}$ and
$70^{+30}_{-40}$ Magellanic satellites in the magnitude range $-7 <
M_{\rm{V}} < -1$ (at 68\% confidence).

The extrapolated contribution of the DES satellites (a total of 250
satellites) is represented by the black diamond in Fig.~\ref{MWsats}.
CDM is consistent with this number particularly for the larger assumed
values of the mass of the Milky Way halo. On the other hand, the
`coldest' 7 keV sterile neutrino, namely $L_6 = 8$, is only marginally
consistent with the extrapolation, while the $L_6 = 12$ and $L_6=700$
models are in significant disagreement with the extrapolated number
count. The predicted number of faint dwarfs produced by any of these
models is, of course, sensitive to the details of the SNfb but in the
following section we consider a limiting case.

\subsection{Model independent constraints on dark matter}
\label{modelindep}

\begin{figure} \centering \includegraphics[trim=0.4cm 0.2cm 0.4cm
  0.2cm,clip=true,width=0.45\textwidth]{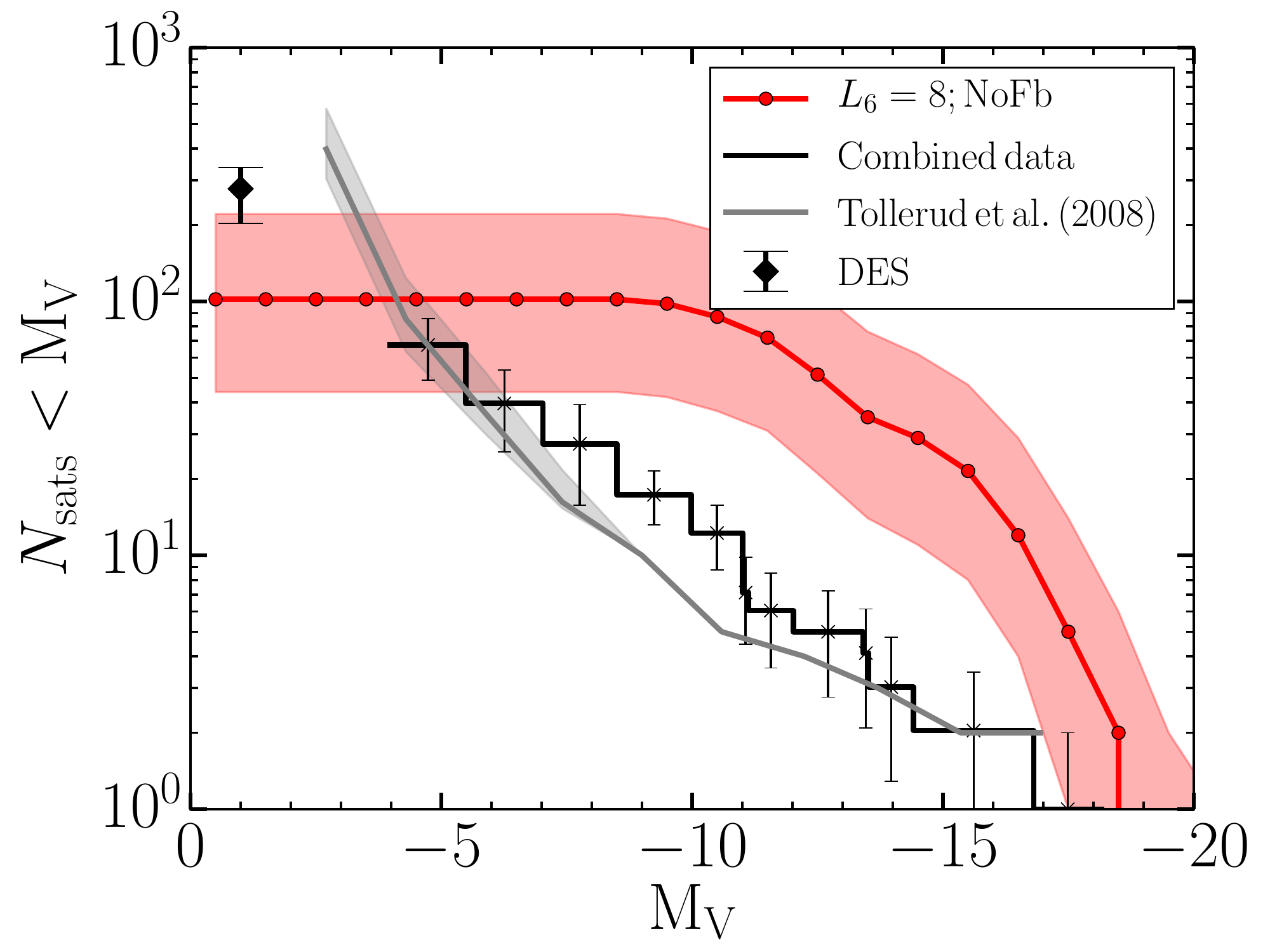}
  \caption{Same as Fig.~\ref{MWsats} for the $L_6=8$ model, but in an extreme 
  scenario where feedback has been turned off completely.}
\label{L68Nofb}
\end{figure}

As mentioned in Section~\ref{sats_of_mw} our analysis suffers from a
degeneracy between the shape of the initial power spectrum and the
strength of SNfb.  A model independent constraint, however, can be
derived by assuming that there is no SNfb at all. In this case, every
subhalo in which gas can cool hosts a satellite, thus maximising the
size of the population. In Fig.~\ref{L68Nofb} we show the predicted
Milky Way satellite luminosity function in the case of zero feedback
(`NoFb').  The total number of satellites is determined entirely by
reionisation i.e., by the amount of gas cooling in haloes prior to the
onset of reionisation.

In Fig.~\ref{L68Nofb} we have assumed $z_{\rm{reion}}^{\rm{full}} =
7.02$, as predicted by the EvoFb scheme for the $L_6=8$ model. This
produces, on average, $\sim 100$ satellites with $M_{\rm{V}} \leq
-1$. A fully self-consistent treatment of reionisation for the NoFb
model would result in $z_{\rm{reion}}^{\rm{full}} > 7.02$, in which
case the number of satellites produced would be even less than 100.
The maximum number of satellite galaxies produced in
Fig.~\ref{L68NoFb} is converged with respect to the halo mass
resolution.  The figure shows that the extreme NoFb model is only
marginally consistent with the extrapolated DES data for the $L_6=8$
case. We recall that this value of the lepton asymmetry corresponds to
the `coldest' possible 7 keV sterile neutrino; ruling this out would
rule out the entire family of 7 keV sterile neutrinos as the dark
matter particles.

The exact location of the extrapolated DES data point in the cumulative
luminosity function is subject to a number of caveats, such as the
DES selection function, detection efficiency and assumptions about
isotropy. However, it is clear that the discovery of even more
ultra-faint dwarf galaxies could potentially set very strong
constraints on the nature of the dark matter.

\section{Conclusions}
\label{conclusions}

We have carried out a detailed investigation of the process of
reionisation in models in which the dark matter particles are assumed
to be sterile neutrinos. The free streaming of these particles leads
to a sharp cut-off in the primordial matter power spectrum at the
scale of dwarf galaxies (Section~\ref{sterile},
Fig.~\ref{ps_models}). On scales much larger than the cut-off,
structure formation proceeds almost identically to CDM. Near and below
the cut-off, sterile neutrinos behave like warm dark matter (WDM): the
abundance of haloes (and therefore of the galaxies they host) is
suppressed and their formation times are delayed relative to CDM. The
sterile neutrino models we consider are motivated by observations of
an X-ray excess at 3.5 keV in the stacked spectrum of galaxy clusters
\citep{Bulbul2014a} and in the spectra of M31 and the Perseus cluster
\citep{Boyarsky2014a}. This excess could be explained by the decay of
a sterile neutrino with a rest mass of 7 keV.

In addition to their rest mass, sterile neutrinos are characterised by
two additional parameters: the lepton asymmetry, $L_6$, and the mixing
angle. Keeping the mass of the sterile neutrino fixed at 7 keV, we
consider three values of $L_6$: 8, 12, 700.  Based on their cut-off
scales, the $L_6=8$ and $L_6=12$ models respectively correspond to the
`coldest' and `warmest' 7 keV sterile neutrinos that are also
consistent with the \cite{Bulbul2014a} and \cite{Boyarsky2014a}
observations. The most extreme model we consider, $L_6=700$, also
decays at 3.5 keV but the mixing angle is unable to produce a decay
flux compatible with the 3.5 keV X-ray observations (see
Table~\ref{dmprops} for a summary).

To calculate the number of ionising photons produced in CDM and in the
sterile neutrino models, we make use of the Durham semi-analytic model
of galaxy formation, \galform~ using the supernova feedback
prescription of \cite{Hou2015}. In this model, the parameters
controlling the strength and evolution of supernova feedback are
calibrated for CDM by the epoch of reionisation as measured by {\it
  Planck}, and tested against data for the luminosity function and
stellar mass-metallicity relation of Milky Way satellites
(Section~\ref{galjun}). We adopt similar values of the model
parameters for our sterile neutrino models. Our main conclusions are:

(i) Although reionisation occurs slightly later in the sterile neutrino models
than in CDM, the epoch of reionisation in all cases is consistent with
the bounds from {\it Planck} (Section~\ref{z_reion},
Fig.~\ref{reion}). For the $L_6=\left(12,700\right)$ models, the
redshifts at which the Universe is 50\% ionised are just below the
68\% confidence interval from {\it Planck}. Reionisation in the
$L_6=8$ model occurs well within the {\it Planck} limits.

(ii) The galaxies that account for the bulk of the ionising photon
budget are more massive in sterile neutrino models than in CDM
(Section~\ref{sources_reion}, Fig.~\ref{sources}).  By the time
reionisation is complete, 50\% of the photoionising budget is produced
by $M_\star \la 10^8\,M_\odot$ galaxies in CDM; the median stellar
mass is $M_\star \sim 10^9\,M_\odot$ for the sterile neutrino models.

(iii) From the evolution of the far-UV luminosity function, we infer
that the galaxy population at high redshift ($z>7$) builds up more
rapidly in the sterile neutrino models than in CDM
(Section~\ref{sources_reion}, Fig.~\ref{uvlfz}). This is particularly
pronounced in the case of the most extreme model, $L_6=700$, which
produces far fewer galaxies than CDM at $z=10$ but `catches up' with
the CDM UV luminosity function by $z=7$. This is directly related to
the more rapid mass accretion of haloes near the free streaming scale
in WDM than in CDM. The qualitative difference in the growth of high
redshift galaxies between CDM and WDM models does not depend on the
details of the galaxy formation model.

(iv) CDM, as well as the three sterile neutrino models we have
considered, are in good agreement with the present-day luminosity
function of the ``classical'' and SDSS Milky Way satellite galaxies
(Section~\ref{sats_of_mw}, Fig.~\ref{MWsats}).  For larger values of
the mass of the Milky Way halo ($M_{\rm{halo}}^{\rm{host}} > 1 \times
10^{12} M_\odot$), even the $L_6=700$ model is consistent with the
observations of \cite{Koposov2008} and \cite{McConnachie2012}.  On the
other hand, if $M_{\rm{halo}}^{\rm{host}} \leq 7 \times 10^{11}
M_\odot$, both the $L_6=700$ and $L_6=12$ models can be ruled out.

(v) Extrapolating to the whole sky the abundance of ultra-faint Milky
Way dwarf satellite galaxies recently detected by DES extends that
satellite luminosity function to very faint magnitudes. With this
extrapolation, the sheer number of satellites places strong
constraints on the sterile neutrino models which produce only a
limited number of substructures. CDM is consistent with this
extrapolation, but the `coldest' 7 keV sterile neutrino (the $L_6 = 8$
model) is only marginally in agreement even when feedback is turned
off completely, a limiting model in which the satellite population is
maximised. Ruling out the $L_6=8$ model, the coolest of the 7keV
sterile neutrino family, would rule out this entire class as
candidates for the dark matter. However, extrapolating the DES counts
to infer the total number of satellites is still subject to a number
of assumptions and uncertainties.

The largest observable differences between CDM and sterile neutrino
models occur at the scale of ultra-faint dwarfs and galaxies at high
redshift. However, only limited data are currently available in these
regimes. The gravitational lensing techniques pioneered by
\cite{Koopmans2005} and \cite{Vegetti2009} may be used to constrain
the subhalo mass function directly, potentially distinguishing WDM
from CDM \citep{RanLi2015}. By increasing the sample of strong lensing
systems, upcoming telescopes such as the SKA and LSST could play a
major role in constraining the nature of the dark matter.

\section*{Acknowledgements}

We would like to thank Mikko Laine, Alexey Boyarsky and Oleg
Ruchayskiy for providing us with the codes for generating the sterile
neutrino power spectra. SB is supported by STFC through grant
ST/K501979/1. This work was supported in part by ERC Advanced
Investigator grant COSMIWAY [grant number GA267291] and the Science
and Technology Facilities Council [grant number ST/F001166/1,
  ST/I00162X/1]. This work used the DiRAC Data Centric system at
Durham University, operated by the Institute for Computational
Cosmology on behalf of the STFC DiRAC HPC Facility
(\url{www.dirac.ac.uk}). This equipment was funded by BIS National
E-infrastructure capital grant ST/K00042X/1, STFC capital grant
ST/H008519/1, and STFC DiRAC Operations grant ST/K003267/1 and Durham
University. DiRAC is part of the National E-Infrastructure. This
research was carried out with the support of the HPC Infrastructure
for Grand Challenges of Science and Engineering Project, co-financed
by the European Regional Development Fund under the Innovative Economy
Operational Programme. The data analysed in this paper can be made
available upon request to the author.

\bibliographystyle{mnras}
\bibliography{sterile.bib}{}

\label{lastpage}

\end{document}